\documentclass[
  aps,prd,preprint,longbibliography,
  showpacs,showkeys,lengthcheck,
  nofootinbib,tightenlines,onecolumn,notitlepage,
  preprintnumbers,superscriptaddress
]{revtex4-2}

\usepackage[utf8]{inputenc}
\usepackage{newtxtext,newtxmath}
\usepackage[mathcal]{euscript}

\usepackage{graphicx}
\usepackage[dvipsnames]{xcolor}
\graphicspath{{figures/}}
\usepackage{tikz-feynman}
\usetikzlibrary{matrix,backgrounds,shadows,shadows.blur}
\usetikzlibrary{decorations.text}
\usetikzlibrary{math}

\usepackage{hyperref}
\hypersetup{colorlinks=true,citecolor=blue,linkcolor=blue,urlcolor=blue}
\usepackage{orcidlink}

\usepackage{diagbox}
\usepackage{booktabs}
\usepackage{siunitx}
\AtBeginDocument{%
  \heavyrulewidth=.08em
  \lightrulewidth=.05em
  \cmidrulewidth=.03em
  \belowrulesep=.65ex
  \belowbottomsep=0pt
  \aboverulesep=.4ex
  \abovetopsep=0pt
  \cmidrulesep=\doublerulesep
  \cmidrulekern=.5em
  \defaultaddspace=.5em
}

\usepackage{amsmath}
\usepackage{slashed}
\usepackage{bm}
\usepackage{bbm}
\usepackage{mathtools}
\usepackage{tensor}

\usepackage[math]{cellspace}
\usepackage[draft,commentmarkup=uwave]{changes} 

\renewcommand*{\d}{\mathop{}\!\mathrm{d}}
\newcommand*{\e}{\mathop{}\!\mathrm{e}}

\newcommand{\hb}[1]{\hat{\bm{{#1}}}}

\newcommand{\bd}{\bm{\varDelta}}
\newcommand{\dl}{\varDelta}
\newcommand{\lrn}{\overleftrightarrow{\nabla}}

\newcommand{\chibar}{\overline{\chi}}

\newcommand{\self}{\bullet}
\newcommand{\cross}{\times}

\begin{document}

\title{
  Stress in static force fields and the sign of the D-term
}

\author{Adam Freese \orcidlink{0000-0002-0688-4121}}
\email{afreese@jlab.org}
\affiliation{Center for Nuclear Femtography, Southeastern Universities Research Association, Newport News, Virginia 23606, USA}
\affiliation{Theory Center, Jefferson Lab, Newport News, Virginia 23606, USA}

\begin{abstract}
  The sign of the D-term is negative for a variety of hadrons,
  including the proton and pion.
  Speculation that this is the result of a mechanical stability condition
  has persisted despite counterexamples,
  including the positive D-term of the hydrogen atom.
  In this work, I explore how the sign of the D-term is influenced by the
  stress carried by static abelian force fields,
  finding the contributions from spin-even fields to be negative
  and the contributions from spin-odd fields
  (such as the electromagnetic field)
  to be positive.
  These contributions correlate with the signs of the D-terms of the
  fields' respective quanta.
  Since fields with different spin can produce identical potentials---thus
  resulting in equally-stable composite systems with identical wave
  functions---the sign of the D-term has nothing to do with stability.
  Since hadrons are bound by spin-one gluons,
  their negative D-terms appear atypical and require explanation.
  I briefly speculate on how color flux confinement could
  produce a negative hadronic D-term.
\end{abstract}

\preprint{JLAB-THY-26-4973}

\maketitle


\section{Introduction}
\label{sec:intro}

Since the groundbreaking work of Maxim Polyakov~\cite{Polyakov:2002yz},
pressure and stress distributions of hadrons
have become a hot topic,
heralding a gold rush of empirical
extractions~\cite{Kumano:2017lhr,Burkert:2018bqq,Kumericki:2019ddg,Duran:2022xag,CLAS:2026lls,CLAS:2026bis},
lattice QCD computations~\cite{Shanahan:2018nnv,Pefkou:2021fni,Hackett:2023rif,Pefkou:2023okb,Hackett:2023nkr},
effective model calculations~\cite{Mai:2012yc,Freese:2019bhb,Neubelt:2019sou,More:2021stk,Lorce:2022cle,Mamo:2022eui,Cao:2023ohj,Cao:2024rul,Hu:2024edc},
and
reviews~\cite{Polyakov:2018zvc,Burkert:2023wzr,Lorce:2025oot,RuizArriola:2026wkb}.
These distributions have also inspired
discussions about their meaning,
including investigations into their dynamical
origin~\cite{Hudson:2017oul,Hudson:2017xug,Maynard:2024wyi}
and
debates about their interpretations as legitimate relativistic
densities~\cite{Lorce:2018egm,Freese:2021czn,Panteleeva:2021iip,Panteleeva:2022uii,Li:2024vgv}.
Discussions have
even included objections to their
meaningfulness~\cite{Ji:2021mfb,Ji:2025gsq,Ji:2025qax}
and, in turn,
defenses of their meaningfulness
too~\cite{Freese:2024rkr,Lorce:2025oot,Cosyn:2026gyy,Liu:2026pbf}.

Stress distributions are given by the spatial components of the energy-momentum tensor (EMT).
They are formally encoded in momentum space by
EMT form factors (EMT-FFs)\footnote{
  The EMT form factors are also often called
  gravitational form factors,
  because the EMT is the source of gravitation
  in general relativity
  and simple extensions such as Einstein-Cartan theory~\cite{cartan1922generalisation}.
  I try to avoid the name ``gravitational form factor''
  after finding that it confuses both laypeople and researchers
  outside the subfield of mechanical properties of hadrons.
}.
Principally among these is the form factor $D(\dl^2)$%
---often called the D-term---%
which gives an abstract momentum-space description
of the stress tensor for a closed system.
The stress distributions themselves can be reconstructed from the D-term
through Fourier transforms, though how to properly
account for relativistic effects while doing so is
controversial~\cite{Lorce:2018egm,Freese:2021czn,Panteleeva:2021iip,Panteleeva:2022uii,Li:2024vgv,Lorce:2025oot}.

Certain global properties of hadrons can be inferred
by looking at the D-term itself.
The D-term at zero momentum transfer, $D(0)$,
quantifies the $r^2$-weighted moment of the isotropic pressure distribution.
Its sign tells us whether the system's internal pressure
is dominated at large distances by negative tensile stresses
(giving $D(0)<0$)
or positive compressive stresses
(giving $D(0)>0$).
Hadrons such as the proton and pion typically have a negative $D(0)$.
It has long been speculated~\cite{Perevalova:2016dln,Polyakov:2018zvc,Lorce:2018egm}
that a negative $D(0)$ is necessary for mechanical stability,
with the story going that compressive pressures at short distances
are needed to protect a system from collapsing in on itself,
while tensile stress at large distance is needed to hold a system together
against disintegration.

Despite persistent speculation along these lines,
there are known stable systems with a positive $D(0)$.
Elementary particles in quantum electrodynamics
such as the electron~\cite{Berends:1975ah,Metz:2021lqv,Freese:2022jlu}
and photon~\cite{Milton:1977je,Freese:2022ibw}
are among them.
Even more strikingly,
the hydrogen atom in its ground state has a positive
$D(0)$~\cite{Ji:2022exr,Czarnecki:2023yqd,Freese:2024rkr}\footnote{
  As an interesting historical curiosity,
  Feynman calculated the stress distributions in the
  hydrogen atom's ground state as part of his bachelor's
  thesis~\cite{Feynman:1939zz}.
  However, he made an overall sign error and mistakenly concluded
  that the atom exhibited short-distance compression
  and long-distance tension.
  Subsequent calculations by Ji, Yang and Liu~\cite{Ji:2022exr},
  Czarnecki, Liu and Reza~\cite{Czarnecki:2023yqd},
  and myself~\cite{Freese:2024rkr}
  all agree that the ordering is really short-distance
  tension and long-distance compression.
}.
The hydrogen atom in particular is a paradigmatic example of a stable system,
seriously calling the purported stability condition into question.

Since the negativity of $D(0)$ is not a universal stability condition,
a better understanding is needed for why
$D(0)$ is negative for the proton and pion
but positive for the hydrogen atom.
More generally, a deeper understanding is needed for
why $D(0)$ ends up having a particular sign for any particular bound system.
While understanding the D-term of the proton is the ultimate goal,
it may not be the best starting point for such an investigation:
it is a complicated relativistic and field-theoretic system
whose structure cannot yet be derived from first principles.
First steps into understanding the sign of the D-term require
simple test systems.
Once these are understood,
the effects of further complications can be investigated,
one addition at a time.

The present work begins where my prior work
on the hydrogen atom~\cite{Freese:2024rkr} left off.
I once again look at the EMT form factors and stress distributions
of two-body quantum systems bound by abelian force fields.
However, I now consider force fields with even spin in addition to the
spin-odd electrostatic field,
and I also consider the effects of massive fields
(and thus of a Yukawa potential).
The principal finding of this work is that abelian forces
mediated by spin-even fields produce a negative D-term,
while abelian fores mediated by a spin-odd field produce a positive D-term.
Since the proton and pion are bound by spin-one gluon fields,
this suggests their negative D-terms are actually \emph{atypical},
with the explanation likely lying in the non-abelian nature
of the gluonic force.

This work is organized as follows.
I present the basic formalism in Sec.~\ref{sec:basic},
along with results for the particle contributions to the form factors.
Sec.~\ref{sec:field} then derives formulas for the field contributions
to the form factors.
This section contains the principal results of the work.
I present numerical illustrations in Sec.~\ref{sec:numeric},
and provide additional discussion in Sec.~\ref{sec:discuss}.
I lastly provide an outlook in Sec.~\ref{sec:end}.


\section{Basic formalism}
\label{sec:basic}

The systems considered in this work are composite spin-zero systems
made up of two spin-zero particles with masses $m_1$ and $m_2$.
The system is non-relativistic
and bound by an abelian static force field
(such as the electrostatic or Yukawa force).
By static, I mean that the field configuration
depends only on the positions of the particles,
$\bm{q}_1$ and $\bm{q}_2$.
The field being abelian allows fields emanating from the two particles
to be added linearly.

The stress tensor can be broken down into contributions from
each of the particles, as well as a field contribution:
\begin{align}
  \hat{T}^{ij}(\bm{x})
  =
  \hat{T}_1^{ij}(\bm{x})
  +
  \hat{T}_2^{ij}(\bm{x})
  +
  \hat{T}_\phi^{ij}(\bm{x})
  \,,
\end{align}
where $1$, $2$ and $\phi$ respectively label the contributions
from particles $1$ and $2$, and the field.
The Pauli stress tensor~\cite{pauli1933allgemeinen}
is used for the particles:
\begin{align}
  \label{eqn:stress:op}
  \hat{T}_n^{ij}(\bm{x})
  =
  \frac{1}{4m_n}
  \Big[
    \hat{p}_n^i ,
    \big[
      \hat{p}_n^j ,
      \delta^{(3)}(\bm{x}-\hb{q}_n)
      \big]_+
    \Big]_+
  \,,
\end{align}
where $[a,b]_+ = ab + ba$ is the anticommutator
and $n \in \{1,2\}$
The field stress tensor obtains its operator structure
from the particle position operators $\hb{q}_1$ and $\hb{q}_2$,
and because of translation invariance depends only on
the displacement of these from the test point $\bm{x}$:
\begin{align}
  \label{eqn:only}
  \hat{T}_\phi^{ij}(\bm{x})
  =
  T_\phi^{ij}(\bm{x}-\hb{q}_1,\bm{x}-\hb{q}_2)
  \,.
\end{align}

Matrix elements of
a single constituent's contribution to the stress tensor
can be broken down in terms of EMT form factors as
follows:
\begin{align}
  \label{eqn:mff}
  \langle \bm{p}' | \hat{T}_a^{ij}(0) | \bm{p} \rangle
  =
  \frac{P^i P^j}{M}
  A_a(\bd^2)
  +
  \frac{\dl^i \dl^j - \delta^{ij} \bd^2}{4M}
  D_a(\bd^2)
  -
  M \delta^{ij}
  \bar{c}_a(\bd^2)
  \,,
\end{align}
where $a \in \{1,2,\phi\}$ labels the constituent,
$M = m_1 + m_2$ is the total mass,
$\bm{p}$ and $\bm{p}'$ are initial and final barycentric momenta,
$\bm{P} = \frac{1}{2} \big( \bm{p} + \bm{p}' \big)$
is the average momentum, and
$\bd = \bm{p}' - \bm{p}$ is the momentum transfer.
Eq.~(\ref{eqn:mff}) is the most general breakdown permitted by Galilei covariance
of the non-relativistic theory,
and is written to correspond to the standard
relativistic breakdown~\cite{Polyakov:2018zvc}
in the non-relativistic limit.

A form factor without a subscript means all constituents
have been summed over:
\begin{align}
  F(\bd^2)
  =
  F_1(\bd^2)
  +
  F_2(\bd^2)
  +
  F_\phi(\bd^2)
  \,,
\end{align}
where $F$ is any of $A$, $D$ or $\bar{c}$.
The so called ``non-conserved'' form factor $\bar{c}(\bd^2)$
must vanish for a closed system~\cite{Ji:1996ek}:
\begin{align}
  \bar{c}(\bd^2)
  =
  0
  \,.
\end{align}
Of the remaining form factors,
$A(\bd^2)$ obeys an additional sum rule
\begin{align}
  A(0)
  =
  1
  \,,
\end{align}
which is known as the momentum sum rule.


\subsection{Particle form factors}
\label{sec:particle}

The formalism to obtain particle form factors
in non-relativistic quantum mechanics
was laid out already in Ref.~\cite{Freese:2024rkr},
but simpler formulas for the form factors
can be obtained than provided there,
\`a la the Bessel transforms of Ref.~\cite{Cosyn:2026gyy}.
A crucial point is that
$|\bm{p}\rangle$ in the form factor breakdown (\ref{eqn:mff})
refers to a composite system with a particular internal wave function,
$\psi(\bm{r})$, and barycentric momentum $\bm{p}$.
By carefully employing completeness relations and Fourier transforms
one can obtain\footnote{
  See Ref.~\cite{Freese:2024rkr} for a step-by-step derivation.
}:
\begin{align}
  \label{eqn:stress:psi}
  \langle \bm{p}' |
  \hat{T}_n^{ij}(0)
  | \bm{p} \rangle
  &=
  \frac{1}{M}
  \int \d^3 \bm{r} \,
  \psi^*(\bm{r})
  \left(
  \chi_n
  P^i P^j
  -
  \frac{i \eta_n}{2}
  \big(P^i \overleftrightarrow{\nabla}^j + \overleftrightarrow{\nabla}^i P^j\big)
  -
  \frac{ \lrn^i \lrn^j }{4\chi_n}
  \right)
  \psi(\bm{r})
  \,
  \e^{i\eta_n \chibar_n\bd\cdot\bm{r}}
  \,,
\end{align}
where
$f \lrn g = f(\nabla g) - (\nabla f)g$,
and where I have introduced several auxiliary quantities
to make the formula easier to read.
These include a sign
\begin{align}
  \label{eqn:eta}
  \eta_1 = +1
  \,, \qquad
  \eta_2 = -1
\end{align}
arising from the sign in front of $\bm{q}_n$
in $\bm{r} = \bm{q}_1 - \bm{q}_2$,
as well as a mass fraction and its complement:
\begin{align}
  \label{eqn:chi}
  \chi_n
  =
  \frac{m_n}{M}
  \,, \qquad
  \chibar_n
  =
  1
  -
  \chi_n
  \,.
\end{align}

The $A_n(\bd^2)$ form factor can be immediately read off from
Eq.~(\ref{eqn:stress:psi}):
\begin{align}
  A_n(\bd^2)
  =
  \chi_n
  \int \d^3 \bm{r} \,
  \psi^*(\bm{r})
  \psi(\bm{r})
  \,
  \e^{i \eta_n \chibar_n \bd\cdot\bm{r}}
  \,.
\end{align}
For S-wave states, this can be further simplified.
Defining the radial wave function by
\begin{align}
  \psi(\bm{r})
  =
  \frac{u(r)}{\sqrt{4\pi} \, r}
  \,,
\end{align}
we can write $A_n(\bd^2)$ as a Bessel transform:
\begin{align}
  \label{eqn:An}
  A_n(\bd^2)
  =
  \chi_n
  \int_0^\infty \d r \,
  u^2(r)
  j_0\big( \chibar_n \dl r \big)
  \,.
\end{align}
The factor $\chibar_n = 1-\frac{m_n}{M}$
damps the falloff of $A_n(\bd^2)$,
with more damping the closer $m_n$ is to $M$.
A particle that carries more of the composite system's mass
is less likely to be very far from the barycenter,
and accordingly will have a low mass radius%
---and hence a small $A'_n(0)$ relative to $A_n(0)$.

The second term of Eq.~(\ref{eqn:stress:psi})
drops out for S-wave states.
For the third term, it is helpful to note that:
\begin{align}
  \label{eqn:helpful}
  -
  \frac{u(r)}{r}
  \frac{
    \lrn^i \lrn^j
  }{4}
  \frac{u(r)}{r}
  =
  \frac{1}{2r^2}
  \bigg\{
    \delta^{ij}
    \left(
    \frac{u^2(r)}{r^2}
    -
    \frac{u'(r) u(r) }{r}
    \right)
    -
    \hat{r}^i \hat{r}^j
    \left(
    \frac{2 u^2(r)}{r^2}
    -
    \frac{u(r) u'(r)}{r}
    +
    u(r) u''(r)
    -
    \big( u'(r) \big)^2
    \right)
    \bigg\}
  \,.
\end{align}
From here, one can use the projection formulas
\begin{align}
  \label{eqn:projections}
  \begin{split}
    D_{a}(\bd^2)
    &=
    \frac{6 M}{\bd^2}
    Y_2^{ij}(\hat{\dl})
    \langle \bm{p}' | T^{ij}_{a}(\bd) | \bm{p} \rangle
    \\
    \bar{c}_{\cross}(\bd^2)
    &=
    -
    \frac{1}{M}
    \hat{\dl}^i \hat{\dl}^j
    \langle \bm{p}' | T^{ij}_{a}(\bd) | \bm{p} \rangle
  \end{split}
\end{align}
to isolate the form factors,
where
\begin{align}
  Y_2^{ij}(\hat{\dl})
  =
  \hat{\dl}^i \hat{\dl}^j
  -
  \frac{1}{3} \delta^{ij}
\end{align}
is an irreducible harmonic tensor~\cite{Polyakov:2018rew}.

With the additional help of the identity
\begin{align}
  \label{eqn:bessel:exp}
  \e^{i\bm{k}\cdot\bm{r}}
  =
  4\pi
  \sum_{l=0}^\infty
  i^l j_l(kr)
  \sum_{m_l=-l}^l
  Y_{lm}^*(\hat{k})
  Y_{lm}(\hat{r})
  \,,
\end{align}
and a few spherical Bessel function identities,
the particle contributions to the remaining form factors are:
\begin{align}
  \label{eqn:Dn}
  D_n(\bd^2)
  &=
  \frac{2}{ \chi_n \dl^2 }
  \int_0^\infty \d r \,
  \left(
  \frac{ 2 u^2(r) }{r^2}
  -
  \frac{ u(r) u'(r) }{r}
  +
  u(r) u''(r)
  -
  \big(u'(r)\big)^2
  \right)
  j_2\big( \chibar_n \dl r \big)
  \\
  \label{eqn:cn}
  \bar{c}_n(\bd^2)
  &=
  \frac{1}{ 2 \chi_n \chibar_n M^2 \dl}
  \int_0^\infty \d r \,
  \Big(
  u'(r) u''(r)
  -
  u(r)u'''(r)
  \Big)
  j_1\big( \chibar_n \dl r \big)
  \,.
\end{align}
There's a strong similarity to
the formulas for $D_U(\bd^2)$ and $\bar{c}_U(\bd^2)$
in Ref.~\cite{Cosyn:2026gyy}.
Both sets of equations coincide when the results here
are specialized to $m_1=m_2$,
and when the results Ref.~\cite{Cosyn:2026gyy}
are specialized to pointlike nucleons with no D-wave.

The $\bar{c}_n(\bd^2)$ form factor
describes forces felt by subsystems~\cite{Polyakov:2018exb},
and in the case of two-body systems in particular,
encodes the force law binding the system~\cite{Freese:2024rkr}.
By using the Schr\"odinger equation,
Eq.~(\ref{eqn:cn}) can be rewritten to make
the relationship to force more manifest:
\begin{align}
  \bar{c}_n(\bd^2)
  &=
  \frac{1}{M\dl}
  \int_0^\infty \d r \,
  u^2(r)
  \big( -V'(r) \big)
  j_1\big( \chibar_n \dl r \big)
  \,.
\end{align}
This formula has two especially appealing uses.
First, it can be inverted to obtain the
force law~\cite{Freese:2024rkr}:
\begin{align}
  -V'(r)
  =
  \frac{ 1 }{ \chibar_n^3 \, | \psi(r) |^2 }
  \frac{\d}{\d b}
  \left[
    \int_0^\infty \d \dl \,
    \dl^2 \bar{c}_n(\dl^2)
    j_0(\dl b)
    \right]
  \bigg|_{b = \chibar_n r}
  \,.
\end{align}
Second, it can be used with the virial theorem
to show the forward limit $\bar{c}_n(0)$ to be:
\begin{align}
  \label{eqn:virial}
  \bar{c}_n(0)
  =
  \frac{1}{3 \chi_n M^2}
  \int_0^\infty \d r \,
  u(r) u''(r)
  <
  0
  \,,
\end{align}
which is a strictly negative quantity.
The condition $\bar{c}(\bd^2) = 0$
then requires that the force field contribution
to $\bar{c}(\bd^2)$ be positive.


\subsection{Force field form factors}
\label{sec:config}

Since it's the main focus of this work,
I'll lay out the formalism to obtain force field form factors
in more detail.
In this subsection, I show how the matrix element
appearing in Eq.~(\ref{eqn:mff})
can be obtained by taking a Fourier transform and an ensemble average
over classical field configurations.
The main result of this section is Eq.~(\ref{eqn:2step}).
This result will be employed for specific classical field configurations
later in Sec.~\ref{sec:field}.

To start, I use two completeness relations:
\begin{align}
  \mathbbm{1}
  =
  \int \d^3 R
  \int \d^3 r
  \,
  | \bm{R}, \bm{r} \rangle
  \langle \bm{R}, \bm{r} |
\end{align}
where $\bm{R}$ and $\bm{r}$ are the standard barycentric
and relative positions:
\begin{align}
  \bm{R}
  =
  \chi_1 \bm{q}_1
  +
  \chi_2 \bm{q}_2
  \,, \qquad
  \bm{r}
  =
  \bm{q}_1
  -
  \bm{q}_2
  \,,
\end{align}
and where $\chi_n = \frac{m_n}{M}$, as in Eq.~(\ref{eqn:chi}).
Additionally, since $|\bm{p}\rangle$ refers to a composite
system with barycentric momentum $\bm{p}$
and internal wave function $\psi(\bm{r})$,
we have:
\begin{align}
  \langle \bm{R}, \bm{r} | \bm{p} \rangle
  =
  \e^{i\bm{p}\cdot\bm{R}}
  \psi(\bm{r})
  \,.
\end{align}
Thus:
\begin{align}
  \langle \bm{p}' |
  \hat{T}^{ij}_\phi(\bm{x})
  | \bm{p} \rangle
  =
  \int \d^3 R
  \int \d^3 R'
  \int \d^3 r
  \int \d^3 r'
  \,
  \psi^*(\bm{r}')
  \psi(\bm{r})
  \e^{i\bm{R}\cdot\bm{p}}
  \e^{-i\bm{R}'\cdot\bm{p}'}
  \langle \bm{R}', \bm{r}' |
  \hat{T}^{ij}_\phi(\bm{x})
  | \bm{R}, \bm{r} \rangle
  \,.
\end{align}
Now, since the field stress depends only on the particle positions
$\bm{q}_1$ and $\bm{q}_2$, and not the particle momenta,
we can treat it as a c-number in the position representation.
With the additional help of Eq.~(\ref{eqn:only}),
this means we can write:
\begin{align}
  \label{eqn:sofar}
  \langle \bm{p}' |
  \hat{T}^{ij}_\phi(\bm{x})
  | \bm{p} \rangle
  =
  \int \d^3 r \,
  \big| \psi(\bm{r}) \big|^2
  \int \d^3 R \,
  \e^{-i\bd\cdot\bm{R}}
  T^{ij}_\phi(\bm{x}-\bm{q}_1, \bm{x}-\bm{q}_2)
  \,.
\end{align}

\begin{figure}
  \includegraphics[scale=1]{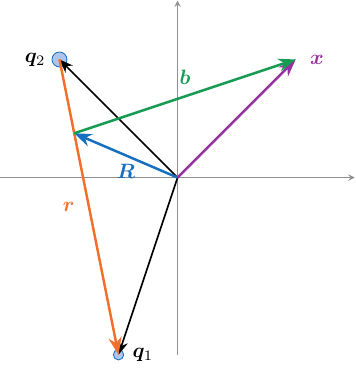}
  \caption{
    Visual depiction of the relationship between
    the particle coordinates $\bm{q}_1$ and $\bm{q}_2$,
    the barycenter $\bm{R}$,
    the relative separation $\bm{r} = \bm{q}_1 - \bm{q}_2$,
    the test point $\bm{x}$,
    and the test point displacement $\bm{b} = \bm{x} - \bm{R}$.
  }
  \label{fig:coordinates}
\end{figure}

It's helpful to consider a more fully specified set of coordinates now.
The coordinates of interest are depicted in Fig.~\ref{fig:coordinates}.
In addition to the barycenter and relative separation,
there is a test coordinate $\bm{x}$ at which the field stress is probed,
and a displacement
\begin{align}
  \bm{b}
  =
  \bm{x}
  -
  \bm{R}
\end{align}
of the test point from the barycenter.
The notation $\bm{b}$ is borrowed from the impact parameter
often appearing in studies of light front
densities~\cite{Burkardt:2000za,Burkardt:2002hr},
since it there has a clear interpretation as a displacement from the barycenter,
and since I am already using $\bm{r}$ for the relative separation.

In Eq.~(\ref{eqn:sofar}),
the field stress depends on displacements from the particle positions.
Since
\begin{align}
  \bm{q}_n
  =
  \bm{R}
  +
  \eta_n \chibar_n \bm{r}
  \,,
\end{align}
with $\chibar_n = 1 - \chi_n$ as in Eq.~(\ref{eqn:chi}),
and where $\eta_n$ is defined in Eq.~(\ref{eqn:eta}),
these displacements can be rewritten:
\begin{align}
  \bm{x}
  -
  \bm{q}_n
  =
  \bm{b}
  -
  \eta_n \chibar_n \bm{r}
  \,.
\end{align}
Changing the integration variable of Eq.~(\ref{eqn:sofar}) to $\bm{b}$ then gives:
\begin{align}
  \label{eqn:2step}
  \langle \bm{p}' |
  \hat{T}^{ij}_\phi(\bm{x})
  | \bm{p} \rangle
  =
  \e^{-i\bm{x}\cdot\bd}
  \int \d^3 r \,
  \big| \psi(\bm{r}) \big|^2
  \int \d^3 b \,
  \e^{i\bd\cdot\bm{b}}
  T^{ij}_\phi\big(\bm{b}-\eta_1\chibar_1\bm{r},\bm{b}-\eta_2\chibar_2\bm{r}\big)
  \,.
\end{align}
We can then evaluate this at $\bm{x}=0$
to find the form factors.

An immediate consequence of Eq.~(\ref{eqn:2step})
is that $A_\phi(\bd^2) = 0$,
since the right-hand side has no $\bm{P}$ dependence.
The remaining form factors can additionally be found
by performing two operations in turn.
First, we can take the Fourier transform of a classical field configuration:
\begin{align}
  \label{eqn:step1}
  \widetilde{T}^{ij}_\phi(\bd;\bm{r})
  =
  \int \d^3 b \,
  \e^{i\bd\cdot\bm{b}}
  T^{ij}_\phi\big(\bm{b}-\eta_1\chibar_1\bm{r},\bm{b}-\eta_2\chibar_2\bm{r}\big)
  \,,
\end{align}
where the integrated variable is the test point displacement $\bm{b}$%
---\emph{not} the configuration variable $\bm{r}$.
Second, we can average this Fourier transform over all possible configurations,
weighted by the square of the configuration space wave function:
\begin{align}
  \label{eqn:step2}
  \langle \bm{p}' |
  \hat{T}^{ij}_\phi(0)
  | \bm{p} \rangle
  =
  \int \d^3 r \,
  \big| \psi(\bm{r}) \big|^2
  \widetilde{T}^{ij}_\phi(\bd;\bm{r})
  \,.
\end{align}
I will employ this two-step procedure in the next section of the work.


\section{Form factors for static force fields}
\label{sec:field}

The focus of this work is forces mediated by abelian fields.
The field configuration will accordingly be the sum of configurations
emanating from each source individually.
I will consider spin-zero, spin-one and spin-two fields.

Let's consider the spin-zero boson fields first.
The following will apply equally to scalar and pseudoscalar fields.
The field Lagrangian is:
\begin{align}
  \mathscr{L}_{\mathrm{scalar}}
  =
  \frac{1}{2} (\partial \phi)^2
  -
  \frac{1}{2}
  \mu^2 \phi^2
  \,,
\end{align}
which entails the following canonical EMT:
\begin{align}
  T^{\mu\nu}_{\mathrm{scalar}}
  =
  (\partial^\mu \phi) (\partial^\nu \phi)
  -
  \frac{1}{2} g^{\mu\nu}
  \big(
  (\partial \phi)^2 - \mu^2 \phi^2
  \big)
  \,.
\end{align}
For a static field (i.e., without time dependence), the stress tensor is:
\begin{align}
  \label{eqn:Tij:scalar}
  T^{ij}_{\mathrm{scalar}}
  =
  (\nabla_i \phi) (\nabla_j \phi)
  -
  \frac{1}{2} \delta^{ij}
  \big(
  (\bm{\nabla} \phi)^2 + \mu^2 \phi^2
  \big)
  \,.
\end{align}
The field emanating from a point source at a position $\bm{q}_n$
with charge $g_n$ is:
\begin{align}
  \label{eqn:phin}
  \phi_n(\bm{x})
  =
  \frac{g_n}{4\pi}
  \frac{\e^{-\mu |\bm{x}-\bm{q}_n|}}{|\bm{x}-\bm{q}_n|}
  \,,
\end{align}
and the field from two point sources at $\bm{q}_1$ and $\bm{q}_2$
with charges $g_1$ and $g_2$
is just the sum:
\begin{align}
  \label{eqn:field:scalar}
  \phi(\bm{x})
  =
  \phi_1(\bm{x})
  +
  \phi_2(\bm{x})
  \,.
\end{align}

Let us next consider spin-one fields.
The following will apply equally to vector and axial vector fields.
The field Lagrangian is:
\begin{align}
  \mathscr{L}_{\mathrm{vector}}
  =
  -
  F^2
  +
  \frac{1}{2}
  \mu^2 A^2
  \,,
\end{align}
where $A$ is the four-vector potential,
$F = \d \wedge A$ is the field strength tensor,
and $\mu$ is the mass of the field.
This Lagrangian applies also to the electromagnetic field when $\mu=0$,
but it is helpful to consider $\mu\neq0$ in intermediate calculations
to regulate infrared divergences.
The energy-momentum tensor can be obtained from Noether's second theorem
using local translations~\cite{Felsager:1981iy,Montesinos:2006th,Freese:2025glz}
or from Belinfante improvement of the canonical
EMT~\cite{Embacher:1986pt,Leader:2013jra}:
\begin{align}
  T^{\mu\nu}_{\mathrm{vector}}
  =
  F^{\mu\rho}
  \tensor{F}{_\rho^\nu}
  +
  \mu^2 A^\mu A^\nu
  +
  \frac{1}{4}
  g^{\mu\nu}
  F^2
  -
  \frac{1}{2} g^{\mu\nu} \mu^2 A^2
  \,.
\end{align}
For static sources in the Coulomb gauge,
$\bm{A} = 0$ and $\bm{B} = 0$.
There is only a non-zero scalar potential
$\phi \equiv A^0$ and electric field $\bm{E} = -\bm{\nabla} \phi$.
The stress tensor in this static scenario is:
\begin{align}
  \label{eqn:Tij:vector}
  T^{ij}_{\mathrm{vector}}
  =
  - E^i E^j
  +
  \frac{1}{2} \delta^{ij}
  \big(
  \bm{E}^2
  +
  \mu^2 \phi^2
  \big)
  =
  -
  (\nabla_i \phi) (\nabla_j \phi)
  +
  \frac{1}{2} \delta^{ij}
  \big(
  (\bm{\nabla} \phi)^2 + \mu^2 \phi^2
  \big)
  \,.
\end{align}
Remarkably, this is the same as Eq.~(\ref{eqn:Tij:scalar}),
except for an overall minus sign.
This extra minus sign arose because the kinetic and mass terms
in the vector field Lagrangian have the opposite signs from
those in the scalar field Lagrangian.
Additionally, the field itself obeys Eq.~(\ref{eqn:field:scalar})
with $g_n = e_n$.

Lastly, let's consider two particles bound by
a static gravitational field.
Strictly speaking, the gravitational field does not contribute
to the covariant energy-momentum tensor in general relativity.
However, it can be helpful in some cases to think of gravitation
in terms of a force field carried by spin-two exchange bosons
which carry energy and momentum~\cite{Butcher:2010ja}.
A variety of energy-momentum pseudotensors exist to this purpose;
see for instance Refs.~\cite{landau1975classical,Butcher:2010ja,Baker:2021qqi,Taylor:2024vvj}.
In the present context---non-relativistic systems bound by static
force fields---it is sufficient to consider Newtonian gravitation.
The relevant Lagrangian is:
\begin{align}
  \mathscr{L}_{\mathrm{gravity}}
  =
  \frac{1}{8\pi G} (\partial \phi)^2
  \,,
\end{align}
where $\phi$ is the gravitational potential and $G$ is Newton's constant.
This Lagrangian is identical to the scalar field Lagrangian with $\mu=0$
up to a factor $\frac{1}{4\pi G}$.
For convenience, this factor can be absorbed into the definition of the field.
The potential will then take the same form as Eq.~(\ref{eqn:field:scalar}),
but with $g_n = \sqrt{4\pi G} m_n$.

In all the relevant cases cases,
the stress tensor can be written in the following form:
\begin{align}
  \label{eqn:Tij:unified}
  \begin{split}
    T^{ij}
    &=
    T^{ij}_{\self,1}
    +
    T^{ij}_{\self,2}
    +
    T^{ij}_{\cross}
    \\
    T^{ij}_{\self,n}
    &=
    (-1)^s
    \left\{
      \big(
      \nabla_i
      \phi_n
      \big)
      \big(
      \nabla_j
      \phi_n
      \big)
      -
      \frac{1}{2} \delta^{ij}
      \big(
      \bm{\nabla}
      \phi_n
      \big)^2
      -
      \frac{1}{2} \delta^{ij}
      \mu^2
      \phi_n^2
      \right\}
    \\
    T^{ij}_{\cross}
    &=
    (-1)^{s}
    \Big\{
      \big(
      \nabla_i
      \phi_1
      \big)
      \big(
      \nabla_j
      \phi_2
      \big)
      +
      (i\leftrightarrow j)
      -
      \delta^{ij}
      \big(
      \bm{\nabla}
      \phi_1
      \big)
      \cdot
      \big(
      \bm{\nabla}
      \phi_2
      \big)
      -
      \delta^{ij}
      \mu^2
      \phi_1
      \phi_2
      \Big\}
    \,.
  \end{split}
\end{align}
Here, the subscripts $\self$ and $\cross$
respectively signify that these are self-field
and interference (or cross-field) stresses,
and $s$ signifies the spin of the force field.

With the unified formula (\ref{eqn:Tij:unified}) for static field stress in hand,
I will now follow the strategy laid out at the end of Sec.~\ref{sec:basic},
by first obtaining the per-configuration Fourier transform
with respect to $\bm{b}$%
---as in Eq.~(\ref{eqn:step1})---%
and then taking the ensemble average over configurations%
---as in Eq.~(\ref{eqn:step2}).
To this end,
it is helpful to rewrite the $\phi_n$ of Eq.~(\ref{eqn:phin})
in terms of its Fourier transform:
\begin{align}
  \phi_n(\bm{b})
  =
  \int \frac{\d^3k}{(2\pi)^3}
  \frac{
    g_n
  }{
    \bm{k}^2 + \mu^2
  }
  \e^{-i\bm{k}\cdot(\bm{x} - \bm{q}_n)}
  \equiv
  \int \frac{\d^3k}{(2\pi)^3}
  \widetilde{\phi}_n(\bm{k})
  \e^{-i\bm{k}\cdot\big(\bm{b} - \eta_n \chibar_n \bm{r}\big)}
  \,,
\end{align}
where $\eta_n$ and $\chibar_n$ are defined in
Eqs.~(\ref{eqn:eta}) and (\ref{eqn:chi}).
The convolution theorem then allows the Fourier transform of
$\phi_1 \phi_2$
to be written as a convolution
$\widetilde{\phi}_1 \otimes \widetilde{\phi}_2$.
The per-configuration Fourier transforms can thus be written:
\begin{align}
  \label{eqn:split}
  \begin{split}
    \widetilde{T}^{ij}_{\self,n}(\bd;\bm{r})
    &=
    (-1)^s
    g_n^2
    \e^{ i \eta_n \chibar_n \bm{\dl}\cdot\bm{r} }
    \int \frac{\d^3 k}{(2\pi)^3}
    \frac{
      k^i k^j
      -
      \frac{1}{4} \dl^i \dl^j
      +
      \frac{1}{2} \delta^{ij}
      \left(
      -
      \bm{k}^2
      +
      \frac{1}{4} \bd^2
      -
      \mu^2
      \right)
    }{
      \left[
        \left(\bm{k} + \frac{1}{2} \bd\right)^2
        +
        \mu^2
        \right]
      \left[
        \left(\bm{k} - \frac{1}{2} \bd\right)^2
        +
        \mu^2
        \right]
    }
    \\
    \widetilde{T}^{ij}_{\cross}(\bd;\bm{r})
    &=
    (-1)^{s}
    g_1 g_2
    \e^{ i \frac{1}{2}\delta \bm{\dl}\cdot\bm{r} }
    \int \frac{\d^3 k}{(2\pi)^3}
    \frac{
      k^i k^j
      -
      \frac{1}{4} \dl^i \dl^j
      +
      \frac{1}{2} \delta^{ij}
      \left(
      -
      \bm{k}^2
      +
      \frac{1}{4} \bd^2
      -
      \mu^2
      \right)
    }{
      \left[
        \left(\bm{k} + \frac{1}{2} \bd\right)^2
        +
        \mu^2
        \right]
      \left[
        \left(\bm{k} - \frac{1}{2} \bd\right)^2
        +
        \mu^2
        \right]
    }
    \e^{i\bm{k}\cdot\bm{r}}
    +
    \mathrm{c.c.}
    \,,
  \end{split}
\end{align}
where
\begin{align}
  \delta
  =
  \chi_2 - \chi_1
  =
  \frac{m_2 - m_n}{M}
\end{align}
is the mass fraction difference.


\subsection{Self-stress term}

Let's continue working out the self-stress.
Starting from the first equation of (\ref{eqn:split}),
Feynman parameters can be used to rewrite the integral as:
\begin{align}
  \label{eqn:divergent}
  \widetilde{T}^{ij}_{\self,n}(\bd;\bm{r})
  &=
  (-1)^s
  g_n^2
  \e^{i\eta_n \chibar_n \bd\cdot\bm{r}}
  \int_0^1 \d y
  \int \frac{\d^3 k}{(2\pi)^3}
  \frac{
    k^i k^j
    -
    \frac{1}{4} (1-y^2) \dl^i \dl^j
    +
    \frac{1}{2} \delta^{ij}
    \left(
    -
    \bm{k}^2
    +
    \frac{1}{4} (1-y^2) \bd^2
    -
    \mu^2
    \right)
  }{
    \left[
      \bm{k}^2
      +
      \frac{1}{4} (1-y^2) \bd^2
      +
      \mu^2
      \right]^2
  }
  \,.
\end{align}
By symmetry, we can also substitute
$k^i k^j \rightarrow \frac{1}{3} \delta^{ij} \bm{k^2}$
in the integrand.
Now, we run into an issue:
the integral is divergent.
The superficial degree of divergence is $3 + 2 - 4 = 1$.
However, the divergence can be regulated
and the integral assigned a finite value through analytic continuation.
One can show that
\begin{align}
  \label{eqn:regulate}
  \int \frac{\d^3 k}{(2\pi)^3}
  \frac{ (\bm{k}^2)^n }{ (\bm{k}^2 + W)^p }
  =
  \frac{W^{3/2 + n - p}}{4\pi^2}
  \frac{
    \Gamma\left(n+\frac{3}{2}\right) \Gamma\left(p-n-\frac{3}{2}\right)
  }{
    \Gamma(p)
  }
\end{align}
whenever $2p - 2n - 3 > 0$.
The right-hand side is finite and analytic in $n$ and $p$,
except for poles in the gamma function
when
$n + \frac{3}{2}$ or $p - n - \frac{3}{2}$
is a non-positive integer.
Aside from these poles,
Eq.~(\ref{eqn:regulate}) can be used as a prescription
to assign finite values to the integral on the left-hand side,
even when the latter diverges\footnote{
  Regulating divergences like this is a much older tradition
  than the renormalization program of quantum field theory.
  This method was used at least as early as 1760 by Euler~\cite{Euler1760}
  (see Ref.~\cite{euler2018divergentseries} for an English translation)
  to assign values to divergent series.
}.
Using this prescription
to evaluate the $\bm{k}$ integral in Eq.~(\ref{eqn:divergent})
gives:
\begin{align}
  \label{eqn:finite}
  \widetilde{T}^{ij}_{\self,n}(\bd;\bm{r})
  &=
  (-1)^{s+1}
  \frac{g_n^2}{16\pi \dl}
  \Big(
  \dl^i \dl^j
  -
  \delta^{ij}
  \bd^2
  \Big)
  \e^{i\eta_n \chibar_n \bd\cdot\bm{r}}
  \int_0^1 \d y \,
  \sqrt{ 1-y^2 + \omega}
  \,.
\end{align}
where I use the auxiliary quantity
\begin{align}
  \omega
  \equiv
  \frac{4 \mu^2}{\bd^2}
\end{align}
to make the formula more compact.
Doing the $y$ integral gives:
\begin{align}
  \label{eqn:Tij:self:tilde}
  \widetilde{T}^{ij}_{\self,n}(\bd;\bm{r})
  &=
  (-1)^{s+1}
  \frac{g_n^2}{32\pi \dl}
  \Big(
  \dl^i \dl^j
  -
  \delta^{ij}
  \bd^2
  \Big)
  \left\{
    (1-\omega)
    \mathrm{asin}\left(\frac{1}{\sqrt{1+\omega}}\right)
    +
    \sqrt{\omega}
    \right\}
  \e^{i\eta_n \chibar_n \bd\cdot\bm{r}}
  \,.
\end{align}
Now, something extraordinary has happened:
the entire structure is proportional to $\dl^i \dl^j - \bd^2 \delta^{ij}$.
The self-stress is conserved, and makes no contribution to $\bar{c}(\bd^2)$.
This is especially extraordinary because
the isotropic average of the field's stress tensor%
---see Eqs.~(\ref{eqn:Tij:scalar}) and (\ref{eqn:Tij:vector})---%
is:
\begin{align}
  \label{eqn:p:sign}
  p(\bm{x})
  =
  \frac{1}{3}
  \delta_{ij}
  T^{ij}(\bm{x})
  =
  \frac{(-1)^{s+1}}{6}
  \big(\bm{\nabla}\phi(\bm{x})\big)^2
  \,,
\end{align}
which is either strictly negative (for even-spin fields)
or strictly positive (for odd-spin).
However, $\bar{c}(0)$ is proportional to the integral of
$p(\bm{x})$ over all space:
\begin{align}
  \int \d^3 x \,
  p(\bm{x})
  =
  -
  M \bar{c}(0)
  \,,
\end{align}
so the condition $\bar{c}(\bd^2)=0$
requires the pressure to flip sign,
in apparent contradiction to Eq.~(\ref{eqn:p:sign}).

The trick happened
when going from the divergent integral in Eq.~(\ref{eqn:divergent})
to the finite result in Eq.~(\ref{eqn:finite}).
Regularizing a strictly positive (or strictly negative) integral or series
can produce a result with the opposite sign~\cite{Euler1760}.
An especially infamous example is the series
$1 + 2 + 3 + 4 + \ldots = -\frac{1}{12}$~\cite{Tao2010emf}.
Another example, pertinent to hadron physics,
is that a renormalized parton distribution function
can become negative~\cite{Collins:2021vke},
despite the bare operator having a probability interpretation.

The sign flip in pressure introduced when regularizing Eq.~(\ref{eqn:divergent})
is not just a mathematical trick,
but has a crucial physical importance.
A closed, stable system needs to satisfy $\bar{c}(\bd^2) = 0$.
For an electrically charged system,
this means negative pressures must be present in addition to the positive
pressure of the system's electrostatic field.
These negative pressures come from Poincar\'e stresses~\cite{Poincare:1906dtc},
which hold the system together against electrostatic self-repulsion.
Especially poignant illustrations of this are shown in recent
classical models of the proton~\cite{Varma:2020crx,Mejia:2025oip}
and electron~\cite{Gardella:2026col}.
The regularization of Eq.~(\ref{eqn:divergent})
implicitly introduces Poincar\'e stresses
(albeit without a detailed mechanistic model),
thus causing $\bar{c}(\bd^2)$ to vanish.

Eq.~(\ref{eqn:Tij:self:tilde}) is general insofar as it can be used to describe
two-body states with any quantum numbers.
At this stage I will use Eq.~(\ref{eqn:step2}) and specialize to S-wave states.
Using the identity (\ref{eqn:bessel:exp})
when performing the angular integration gives:
\begin{align}
  \langle \bm{p}' |
  \hat{T}^{ij}_{\self,n}(0)
  | \bm{p} \rangle
  &=
  (-1)^{s+1}
  \frac{g_n^2}{32\pi \dl}
  \Big(
  \dl^i \dl^j
  -
  \delta^{ij}
  \bd^2
  \Big)
  \left\{
    (1-\omega)
    \mathrm{asin}\left(\frac{1}{\sqrt{1+\omega}}\right)
    +
    \sqrt{\omega}
    \right\}
  \int_0^\infty \d r \,
  u^2(r)
  j_0\big( \chibar_n \dl r \big)
  \,.
\end{align}
It is straightforward to read off the $D(\bd^2)$ form factor from this.
Using a trigonometric identity and restoring the original variables then gives:
\begin{align}
  \label{eqn:D:self}
  D_{\self,n}(\bd^2)
  =
  (-1)^{s+1}
  \frac{g_n^2 M}{8\pi\dl}
  \left\{
    \left(1 - \frac{4 \mu^2}{\dl^2}\right)
    \mathrm{atan}\left(\frac{\dl}{2\mu}\right)
    +
    \frac{ 2\mu }{\dl}
    \right\}
  \int_0^\infty \d r \,
  u^2(r)
  j_0\big( \chibar_n \dl r \big)
  \,.
\end{align}


\subsection{Interference stress term}

The interference stress terms are more difficult to deal with,
owing to the presence of $\e^{\pm i\bm{k}\cdot\bm{r}}$
in the second equation of (\ref{eqn:split}).
Like with the self-stress term,
we can start by using Feynman parametrization:
\begin{multline}
  \widetilde{T}^{ij}_{\cross}(\bd;\bm{r})
  =
  \frac{
    (-1)^{s}
    g_1 g_2
  }{2}
  \e^{i \frac{1}{2}\delta \bd\cdot\bm{r}}
  \int_{-1}^1 \d y \,
  \e^{iy \frac{\bd\cdot\bm{r}}{2}}
  \\
  \int \frac{\d^3 k}{(2\pi)^3}
  \frac{
    k^i k^j
    +
    \frac{y}{2}( k^i \dl^j + \dl^i k^j )
    +
    \frac{1}{4}(1-y^2) \dl^i \dl^j
    +
    \frac{1}{2} \delta^{ij}
    \left(
    -
    \bm{k}^2
    -
    y \bd\cdot\bm{k}
    -
    \frac{1}{4}(1-y^2) \bd^2
    -
    \mu^2
    \right)
  }{
    \left[ \bm{k}^2 + \frac{1}{4}(1-y^2) \bd^2 + \mu^2 \right]^2
  }
  \e^{i\bm{k}\cdot\bm{r}}
  +
  \mathrm{c.c.}
  \,.
\end{multline}
Each $\bm{k}$ can be turned into $-i\bm{\nabla}$ acting on
the complex exponent, giving:
\begin{multline}
  \widetilde{T}^{ij}_{\cross}(\bd;\bm{r})
  =
  \frac{
    (-1)^{s}
    g_1g_2
  }{2}
  \e^{i \frac{1}{2}\delta \bd\cdot\bm{r}}
  \int_{-1}^1 \d y \,
  \e^{iy \frac{\bd\cdot\bm{r}}{2}}
  \bigg\{
    -
    \nabla_i \nabla_j
    -
    \frac{iy}{2}( \nabla^i \dl^j + \dl^i \nabla^j )
    -
    \frac{1}{4}(1-y^2) \dl^i \dl^j
    \\
    +
    \frac{1}{2} \delta^{ij}
    \left(
    \bm{\nabla}^2
    +
    i
    y \bd\cdot\bm{\nabla}
    +
    \frac{1}{4}(1-y^2) \bd^2
    -
    \mu^2
    \right)
    \bigg\}
  \int \frac{\d^3 k}{(2\pi)^3}
  \frac{
    \e^{i\bm{k}\cdot\bm{r}}
  }{
    \left[ \bm{k}^2 + \frac{1}{4}(1-y^2) \bd^2 + \mu^2 \right]^2
  }
  +
  \mathrm{c.c.}
  \,.
\end{multline}
The $\bm{k}$ integral can be done using the identity
\begin{align}
  \int \frac{\d^3 k}{(2\pi)^3}
  \frac{
    \e^{i\bm{k}\cdot\bm{r}}
  }{
    ( \bm{k}^2 + B^2 )^2
  }
  =
  \frac{ \e^{-Br} }{8\pi B}
  \,,
\end{align}
giving:
\begin{multline}
  \widetilde{T}^{ij}_{\cross}(\bd;\bm{r})
  =
  \frac{(-1)^{s+1}g_1 g_2 \dl}{32\pi}
  \e^{i \frac{1}{2}\delta \bd\cdot\bm{r}}
  \int_{-1}^1 \d y \,
  \e^{iy \frac{\bd\cdot\bm{r}}{2}}
  \bigg\{
    \sqrt{1-y^2 + \omega}
    \Big(
      \hat{r}^i \hat{r}^j
      +
      \hat{\dl}^i \hat{\dl}^j
      -
      \delta^{ij}
      \Big)
    +
    \frac{2}{\dl r}
    \hat{r}^i \hat{r}^j
    \\
    -
    iy\big(
    \hat{r}^i \hat{\dl}^j
    + \hat{\dl}^i \hat{r}^j
    - \delta^{ij} \cos\theta_{r\dl}
    \big)
    -
    \frac{
      \omega (\hat{\dl}^i \hat{\dl}^j - \delta^{ij})
    }{\sqrt{1-y^2+\omega}}
    \bigg\}
  \e^{-\frac{\dl r}{2} \sqrt{1 - y^2 + \omega}}
  +
  \mathrm{c.c.}
  \,.
\end{multline}
The expression so far is general in the sense that it can be used for a two-body
system with arbitrary quantum numbers.
It will accordingly be used as a launching point for future work considering
spin-one systems.
For now, however, I will now specialize to the case of an $S$-wave state.
Again using Eqs.~(\ref{eqn:step2}) and (\ref{eqn:bessel:exp})%
---along with some spherical Bessel function identities and integration by parts---%
gives:
\begin{multline}
  \label{eqn:D:cross}
  D_{\cross}(\bd^2)
  =
  (-1)^{s+1}
  \frac{g_1 g_2 M}{4\pi\dl^2}
  \int_0^\infty \d r \,
  u^2(r)
  \Bigg\{
    \frac{ (1-\omega) \dl }{2}
    \,
    \Phi\left(\frac{\dl r}{2}, \omega, \delta\right)
    \\
    +
    \left[
      \frac{6(1+\mu r)}{\dl r}
      \Big(
      j_1\big(\chi_1 \dl r\big)
      +
      j_1\big(\chi_2 \dl r\big)
      \Big)
      -
      \Big(
      j_0\big(\chi_1 \dl r\big)
      +
      j_0\big(\chi_2 \dl r\big)
      \Big)
      \right]
    \frac{ \e^{-\mu r}}{r }
    \Bigg\}
\end{multline}
for the $D$ form factor and
\begin{align}
  \label{eqn:cbar:cross}
  \bar{c}_{\cross}(\bd^2)
  &=
  (-1)^{s}
  \frac{g_1 g_2}{4 \pi M \dl}
  \int_0^\infty \d r \,
  u^2(r)
  \frac{(1+\mu r) \e^{-\mu r} }{r^2}
  \Big(
  j_1\big(\chi_1 \dl r\big)
  +
  j_1\big(\chi_2 \dl r\big)
  \Big)
\end{align}
for $\bar{c}$.
In the result for $D_{\cross}(\bd^2)$,
I used the auxiliary function:
\begin{align}
  \label{eqn:Phi:definition}
  \Phi(\zeta,\omega, \delta)
  &\equiv
  \int_{-1}^1 \d y \,
  \frac{
    \e^{-\zeta\sqrt{1-y^2+\omega}}
  }{\sqrt{1-y^2+\omega}}
  j_0\big( (y+\delta)\zeta \big)
  \,.
\end{align}
It is possible to write $\Phi(\zeta,\omega,\delta)$ in terms of known special
functions---the exponential integral function $E_1$ in particular---but
the expression is not pretty or compact.
The analytic expression (and its derivation)
have accordingly been exiled to Appendix~\ref{sec:functions}.

Formulas for the force field form factors have now been fully specified.
We have:
\begin{align}
  \label{eqn:mff:force}
  \begin{split}
    A_\phi(\bd^2)
    &=
    0
    \\
    D_\phi(\bd^2)
    &=
    D_{\self,1}(\bd^2)
    +
    D_{\self,2}(\bd^2)
    +
    D_{\cross}(\bd^2)
    \\
    \bar{c}_\phi(\bd^2)
    &=
    \bar{c}_{\cross}(\bd^2)
    \,,
  \end{split}
\end{align}
where the individual pieces are given in Eqs.~(\ref{eqn:D:self}),
(\ref{eqn:D:cross}) and (\ref{eqn:cbar:cross}).
The force field does not contribute to $A(\bd^2)$,
the self-fields contribute to $D(\bd^2)$ but not to $\bar{c}(\bd^2)$,
and the interference between the particles' fields
contribute to both $D(\bd^2)$ and $\bar{c}(\bd^2)$.
The vanishing of the self-field contribution to $\bar{c}(\bd^2)$
occurs mathematically from the regularization of a divergent integral
defining the form factor,
and is physically necessary for each particle to separately be a stable system.


\subsection{Massless fields and the forward limit}

The case of massless fields is especially pertinent,
since this includes the electromagnetic and gravitational fields.
It is especially worthwhile to examine the forward limit, $\bd^2=0$,
because of known pathologies involving infinite $D(0)$
for electrically charged
systems~\cite{Metz:2021lqv,Freese:2022jlu,Mejia:2025oip,Gardella:2026col}.

Let's first look at $D_\phi(\bd^2)$.
For the self-field contributions,
we just set $\mu=0$ in Eq.~(\ref{eqn:D:self}):
\begin{align}
  D_{\self,n}(\bd^2)
  =
  (-1)^{s+1}
  \frac{g_n^2 M}{16 \dl}
  \int_0^\infty \d r \,
  u^2(r)
  j_0\big( \chibar_n \dl r \big)
  \,.
\end{align}
This has a pole at $\dl=0$:
\begin{align}
  D_{\self,n}(\bd^2)
  =
  (-1)^{s+1}
  \frac{g_n^2 M}{16 \dl}
  +
  \mathcal{O}(\dl)
  \,.
\end{align}
The interference contribution is found by setting
$\mu = 0$ in Eq.~(\ref{eqn:D:cross}):
\begin{align}
  D_{\cross}(\bd^2)
  =
  (-1)^{s+1}
  \frac{g_1 g_2 M}{8 \pi \dl}
  \int_0^\infty \d r \,
  u^2(r)
  \left\{
    \Phi\left(\frac{\dl r}{2}, 0, \delta\right)
    +
      \frac{
        12
        \big(
        j_1(\chi_1 \dl r)
        +
        j_1(\chi_2 \dl r)
        \big)
      }{
        (\dl r)^2
      }
      -
      \frac{
        2
        \big(
        j_0(\chi_1 \dl r)
        +
        j_0(\chi_2 \dl r)
        \big)
      }{
        \dl r
      }
    \right\}
  \,.
\end{align}
This also has a pole at $\dl=0$.
It is a simple pole;
the Taylor expansions of $j_0$ and $j_1$ guarantee
that the extra negative powers of $\dl$ multiplying
these functions cancel out.
By Taylor-expanding the spherical Bessel functions,
as well as the integrand in Eq.~(\ref{eqn:Phi:definition}),
one can show that:
\begin{align}
  \label{eqn:D0:massless:cross}
  D_{\cross}(\bd^2)
  =
  (-1)^{s+1}
  g_1 g_2 M
  \left[
    \frac{1}{8 \dl}
    -
    \frac{7+\delta^2}{60\pi}
    \int_0^\infty \d r \,
    r u^2(r)
    \right]
  +
  \mathcal{O}(\dl)
  \,.
\end{align}
Adding both contributions gives:
\begin{align}
  \label{eqn:D0:massless}
  D_\phi(\bd^2)
  =
  (-1)^{s+1} M
  \left\{
    \frac{(g_1+g_2)^2}{16\dl}
    -
    \frac{g_1 g_2 (7+\delta^2)}{60\pi}
    \int_0^\infty \d r \, r u^2(r)
    \right\}
  +
  \mathcal{O}(\dl)
  \xrightarrow[\dl\rightarrow0]{}
  \left\{
    \begin{array}{lcl}
      \text{finite} ~~&:&~~ g_1 = -g_2 \\
      -\infty       ~~&:&~~ g_1 \neq -g_2, ~~ \text{spin-odd} \\
      +\infty       ~~&:&~~ g_1 \neq -g_2, ~~ \text{spin-even} \\
    \end{array}
    \right.
  \,.
\end{align}
The potential pole at $\dl=0$ drops out iff $g_1 = -g_2$,
meaning $D_U(0)$ is finite iff the particles
have equal and opposite charges.
In all other cases, $D_U(0)$ will be infinite,
with the sign of the infinity being positive for spin-odd fields
and negative for spin-even.

This finding is consistent with prior results finding
an infinite electromagnetic field contribution to $D(0)$ for the
proton~\cite{Varma:2020crx,Mejia:2025oip}
and
electron~\cite{Metz:2021lqv,Freese:2022jlu,Gardella:2026col}
and this effectively occurs because of the slow falloff of the static field.
This has already been explained meticulously in the pioneering work
of Varma and Schweitzer~\cite{Varma:2020crx},
but I will reiterate the explanation here.
In the far field region,
the electrostatic field of any system with a net electric charge will
go as $r^{-2}$,
meaning the isotropic pressure will go as $r^{-4}$.
Ignoring $\bar{c}(\bd^2)$
(which will be zero for the total system anyway),
$D(0)$ is proportional to the $r^2$ moment of the isotropic pressure:
\begin{align}
  D(0)
  =
  \frac{1}{3M}
  \int \d^3 r \,
  r^2 p(\bm{r})
  \,,
\end{align}
which means the far field region%
---that is, $r \gg R$ for some radius $R$ much larger
than the size of the system---%
will make a contribution
\begin{align*}
  D(0)
  \sim
  \int_R^\infty \d r \,
  r
  =
  \infty
  \,.
\end{align*}
The infinity in $D(0)$ is an infrared divergence
caused by the electrostatic field falling too slowly with distance.

On the other hand, any system with net charge zero will have a finite $D(0)$.
The field in the far-field region cannot fall more slowly than $r^{-3}$%
---and thus the stress as $r^{-6}$---%
so the large-$r$ divergence appearing when $g_1 \neq -g_2$ will not occur.
Hence $D(0)$ for the hydrogen atom for instance is positive,
but finite~\cite{Ji:2022exr,Czarnecki:2023yqd,Freese:2024rkr}.
A crucial point worth noticing here is that the net charge must be zero
for $D(0)$ to be finite:
an $\mathrm{He}^+$ ion will have $D(0) = +\infty$,
despite having the same potential and wave function as the hydrogen atom.

Along the same vein,
the $D(0)$ of any gravitationally bound system
will be negative and infinite.
Gravitation is mediated by a spin-two field,
and all gravitating systems have identically-signed charges%
---the latter just being a consequence of the equivalence principle.
Put another way, all systems have a non-zero net gravitational charge,
meaning that there is a monopole-like $r^{-2}$ falloff in the gravitational field
in the far-field region.
Getting $D(0) = -\infty$ for gravitationally bound systems thus has
the same structural cause as getting $D(0) = +\infty$
for electrostatic systems with non-zero net change,
but the sign is negative because the gravitational field has even spin.
However, this conclusion that $D(0) = -\infty$ can be avoided if the
gravitational field is stipulated not to carry energy or momentum,
as in the standard formulation of general relativity.

Let's consider $\bar{c}(\bd^2)$ next.
There is only an interference contribution,
and we can just set $\mu = 0$ in Eq.~(\ref{eqn:cbar:cross}):
\begin{align}
  \bar{c}_{\phi}(\bd^2)
  &=
  (-1)^{s}
  \frac{g_1 g_2}{4 \pi M \dl}
  \int_0^\infty \d r \,
  \frac{ u^2(r) }{r^2}
  \Big(
  j_1\big(\chi_1 \dl r\big)
  +
  j_1\big(\chi_2 \dl r\big)
  \Big)
\end{align}
This is finite at $\dl=0$:
\begin{align}
  \bar{c}_\phi(0)
  &=
  (-1)^{s}
  \frac{g_1 g_2}{12 \pi M}
  \int_0^\infty \d r \,
  \frac{ u^2(r) }{r}
  \,.
\end{align}
The integrand is positive,
so the sign of $\bar{c}_\phi(0)$ will be the same as the sign
of $(-1)^s g_1 g_2$.
As we saw in Eq.~(\ref{eqn:virial}),
the sign of $\bar{c}_n(0)$ is strictly negative.
In order to have $\bar{c}(0) = 0$,
we must have $\bar{c}_\phi(0) > 0$.
This can only occur for spin-even fields if $g_1$ and $g_2$
have the same sign,
or for spin-odd fields if $g_1$ and $g_2$ have opposite signs.
This is consistent with the rule that spin-one fields produce
attractive forces for opposite charges,
while spin-zero and spin-two fields produce
attractive forces for like charges~\cite{Zee:2003mt}.


\subsection{Forward limit for massive fields}

The forward limit is also worth examining for massive fields.
There are two reasons to do so.
First, even for massless fields like the electrostatic field,
it can be instructive to regulate infrared divergences using a small photon mass
which is ultimately taken to zero.
Second, forces mediated by scalar boson exchange will typically involve
massive scalar bosons and like-charged particles.
Since $D(0)$ is infinite for massless fields unless the particle charges
are exactly opposite,
the scalar field mass will be vital to obtaining a finite $D(0)$.

The self-field contribution to $D(0)$ can be found by
setting $\dl=0$ in Eq.~(\ref{eqn:D:self}):
\begin{align}
  \label{eqn:D0:massive:self}
  D_{\self,n}(0)
  =
  (-1)^{s+1}
  \frac{g_n^2 M}{12 \pi \mu}
  \,.
\end{align}
This diverges in the $\mu \rightarrow 0$ limit,
and is positive for odd-spin fields and negative for even-spin fields.
Curiously, this result is consistent with the small-mass limit
of the electron $D(\bd^2)$ in QED with a massive photon~\cite{Metz:2021lqv}.
This makes sense, since a self-field contribution to the stress tensor
is present for one-particle states.

The interference contribution is straightforward but a bit tedious to obtain.
Eq.~(\ref{eqn:D:cross}) must be Taylor-expanded in $\dl$.
As long as $\mu > 0$, the $\dl^{-1}$ terms cancel among themselves,
and the end result is finite in the $\dl \rightarrow 0$ limit.
Adding the result to the self-field D-term gives:
\begin{align}
  \label{eqn:D0:massive}
  D_\phi(0)
  =
  \frac{ (-1)^{s+1} M }{4\pi}
  \left\{
    \frac{(g_1+g_2)^2}{3\mu}
    +
    \int_0^\infty \d r \,
    r
    u^2(r)
    \e^{-\mu r}
    \left[
      \frac{1}{5}
      \left(1-\frac{\mu r}{9}\right)
      -
      \frac{2}{3}
      \frac{\e^{\mu r}-1}{\mu r}
      -
      \frac{(1+\mu r)\delta^2}{15}
      \right]
    \right\}
  \,.
\end{align}
The term that potentially diverges in the $\mu\rightarrow0$ limit
drops out iff $g_1 = -g_2$,
and the remaining term reduces in this limit to the finite
term in Eq.~(\ref{eqn:D0:massless}).
This validates consistency between the
$\Delta \rightarrow 0$ and $\mu \rightarrow 0$ limits.

Lastly, the forward limit of Eq.~(\ref{eqn:cbar:cross}) is:
\begin{align}
  \bar{c}_\phi(0)
  &=
  (-1)^{s}
  \frac{g_1 g_2}{12 \pi M}
  \int_0^\infty \d r \,
  u^2(r)
  \frac{(1 + \mu r)\e^{-\mu r}}{r}
  \,.
\end{align}
This is finite, and has the same sign as the massless case.


\section{Numerical examples}
\label{sec:numeric}

This section presents a few numerical illustrations of how the spin and mass
of an abelian force field affect the EMT form factors
$D(\bd^2)$ and $\bar{c}(\bd^2)$.
The numerical calculations are done using
deupack~\cite{Freese_deupack_2026}.
Since deupack was designed to calculate EMT form factors of the deuteron,
it assumes equal mass constituents ($m_1 = m_2 \equiv m$),
which I shall in turn assume throughout this section.


\subsection{Massless force field}

The potential energy function for two particles interacting through
massless boson exchange is the Coulomb potential:
\begin{align}
  V(r)
  =
  \frac{(-1)^{s+1} g_1 g_2}{4\pi r}
  \equiv
  -
  \frac{\alpha}{4\pi r}
  \,.
\end{align}
I will assume
$\alpha \equiv (-1)^{s} g_1 g_2 > 0$
so that the potential is attractive,
since we are studying mechanical properties of bound states.
The ground state solution to the Schr\"odinger equation
with this potential is well-known:
\begin{align}
  u(r)
  =
  \sqrt{\frac{(\alpha m)^3}{2}}
  \,
  r
  \e^{-\frac{1}{2}\alpha m r}
  \,,
\end{align}
where $\frac{m}{2}$ is the reduced mass.

EMT form factors for the ground state solution
were already been examined by
Ji, Yang and Liu~\cite{Ji:2022exr}
in the case of infinite mass imbalance,
and later by myself~\cite{Freese:2024rkr} for general masses.
The interesting point to examine here is how the form factors depend
on the spin of the field responsible for the potential.
Thus, in addition to an electrostatically bound system,
I shall consider a gravitationally bound system.

For the electrostatic system, I consider (true) muonium:
a bound state of a muon and anti-muon in its ground state.
Thus $m = m_\mu \approx 105.658$~MeV,
and
$\alpha = \alpha_{\mathrm{QED}} \approx 0.007297352$.
For the gravitational system, I consider a somewhat contrived
system that I shall call keplerium.
The mass $m_k$ of each constituent
is chosen so that the effective Bohr radius is the same as muonium:
\begin{align}
  \label{eqn:mk}
  m_k
  =
  \sqrt[3]{\frac{\alpha_{\mathrm{QED}} m_\mu}{G}}
  \approx
  486.19~\mathrm{EeV}
  \approx
  866.72~\mathrm{pg}
  \,,
\end{align}
where $G \approx 6.70883\cdot10^{-39}~\mathrm{GeV}^{-2}$
is Newton's gravitational constant,
and where EeV is exaelectrocvolts
and pg is picograms.
The mass $m_k$ is peculiar in its enormity relative to standard
high-energy physics scales,
but its tininess relative to macroscopic objects.
Still, it produces the same effective Bohr radius under gravitational attraction
as muonium under electrostatic attraction,
so keplerium and muonium have numerically identical wave functions.

\begin{figure}
  \includegraphics[width=\textwidth]{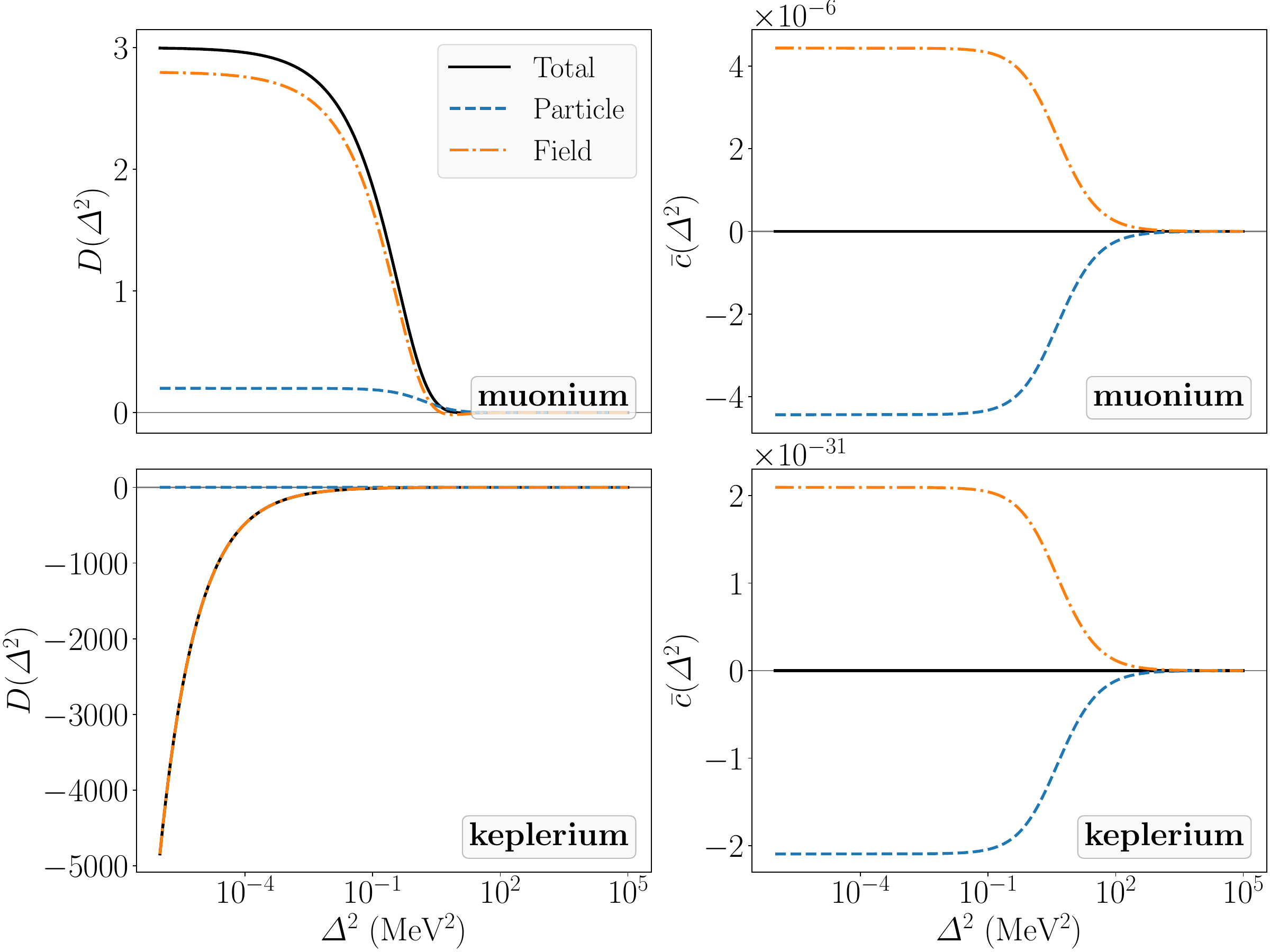}
  \caption{
    EMT form factors for the ground state of muonium (top two panels)
    and keplerium (bottom two panels).
    Muonium is a bound state of muon and anti-muon under electrostatic attraction,
    and keplerium is a bound state of two particles under gravitational attraction
    with masses chosen to produce the same effective Bohr radius as muonium.
    The blue dashed lines are particle contributions to the EMT-FFs,
    the orange dash-dotted lines are static field contributions,
    and the solid black line is the sum.
  }
  \label{fig:massless}
\end{figure}

Despite having the same wave function,
muonium and keplerium do not have identical EMT form factors.
These are shown in Fig.~\ref{fig:massless},
with the muonium form factors in the top two panels
and keplerium in the bottom two.
The first point of comparison is that $\bar{c}_a(\bd^2)$---see the
right two panels---are the same up to an overall scale.
This can be explained by the dependence of $\bar{c}_n(\bd^2)$ on
the system's mass, as in Eq.~(\ref{eqn:cn}).
Keplerium is significantly more massive than muonium,
and $\bar{c}_n(\bd^2)$ is inversely proportional to the square
of the system's mass.
As for $\bar{c}_\phi(\bd^2)$, it is guaranteed by local
momentum conservation to equal $-\bar{c}_1(\bd^2) - \bar{c}_2(\bd^2)$,
so it must scale the same way with system mass as $\bar{c}_n(\bd^2)$.

There is a significant difference, however,
in the $D(\bd^2)$ form factor.
As already shown in prior works~\cite{Ji:2022exr,Czarnecki:2023yqd,Freese:2024rkr},
$D(0)$ for the ground state of an electrostatically bound, neutral particle
is finite and positive.
However, for a gravitationally bound system,
$D(\bd^2)$ is negative and $D(0) = -\infty$.
This was already shown in Eq.~(\ref{eqn:D0:massless}),
but the bottom-left panel of Fig.~\ref{fig:massless} shows this graphically.


\subsection{Massive force field}
\label{sec:numeric:massive}

For massive fields, the potential energy function
is given by the Yukawa potential~\cite{Yukawa:1935xg}:
\begin{align}
  V(r)
  =
  \frac{(-1)^{s+1} g_1 g_2}{4\pi r}
  \e^{-\mu r}
  \equiv
  -
  \frac{\alpha}{r}
  \e^{-\mu r}
  \,.
\end{align}
There are no known analytic solutions
to the Schr\"odinger equation
with this potential.
A numerical estimate of the wave function is required.

Although no analytic solutions are known,
the Yukawa potential has been well-studied.
A variety of precise numerical solutions for bound states
and energy levels exist;
see Ref.~\cite{Edwards:2017ndv} for a review.
I will take a comparatively simple variational approach,
which is sufficient for the purposes of this work,
but it must be informed by several known peculiarities
of the Yukawa potential.

Firstly---unlike the Coulomb potential---the Yukawa potential
has only finitely many solutions.
In fact, for $\mu$ above some critical value~\cite{Edwards:2017ndv}\footnote{
  In Ref.~\cite{Edwards:2017ndv},
  $m$ signifies the reduced mass,
  whereas here $\frac{m}{2}$ is the reduced mass.
  Any apparent factor $2$ discrepancies between this work
  and Ref.~\cite{Edwards:2017ndv}
  are just differences of notation.
}
\begin{align}
  \mu_c
  \approx
  1.190612
  \alpha
  \frac{m}{2}
  \,,
\end{align}
there are no bound states at all.
The values of $m$, $\mu$ and $\alpha$ must be chosen
to ensure the existence of a bound state.
Moreover, as meticulously explored in Ref.~\cite{Edwards:2017ndv},
standard methods for numerical estimates tend to be more accurate
for $\mu \ll \mu_c$.
It is thus prudent to stay within this regime.
For concreteness,
I consider
$m=1~\mathrm{GeV}$,
$\alpha=1$,
and
$\mu \leq 0.1~\mathrm{GeV}$
(with multiple values of $\mu$ to be considered later).

I use a two-step variational method
to approximate the ground state wave function.
The first step obtains an estimate for the ground state energy,
and the second step estimates the wave function.
In both steps,
I use the parametric form
\begin{align}
  \label{eqn:trial}
  u(r)
  =
  r
  (1 + a_1 r + a_2 r^2 + \ldots + a_N r^N)
  \e^{-\kappa r}
  \,.
\end{align}

In the first step,
the coefficients $a_1, \ldots a_N$ and
the exponential decay factor $\kappa$
are all allowed to float.
The expected value of the energy
\begin{align}
  \langle E \rangle
  =
  \frac{1}{\int_0^\infty \d r \, u^2(r)}
  \int_0^\infty \d r \,
  \left(
  -
  \frac{u''(r)}{m}
  +
  V(r) u(r)
  \right)
  u(r)
  \,,
\end{align}
is minimized
through the differential evolution method~\cite{storn1997differential}
in SciPy~\cite{2020SciPy-NMeth}.
It is well-known that solutions for the Yukawa potential
should have a decay constant $\kappa$ given by
\begin{align}
  \label{eqn:kappa}
  \kappa
  =
  \sqrt{-m E}
  \,,
\end{align}
but the solutions found in this first step typically
do not satisfy this relation
because the large-$r$ part of the wave function contributes
little to $\langle E \rangle$.
Thus, as a second step, I repeat the fit with a fixed
$\kappa$ determined by Eq.~(\ref{eqn:kappa}),
allowing only the coefficients $a_1, \ldots, a_N$ to float.

\begin{table}
  \renewcommand{\arraystretch}{2.0}
  \caption{
    Estimates for the ground state energy and wave function
    for the Yukawa potential,
    given two particles with equal mass $m=1~\mathrm{GeV}$,
    a screening parameter $\mu = 0.1~\mathrm{GeV}$,
    and a coupling strength $\alpha = 1$.
    The coefficients are for the
    trial wave function (\ref{eqn:trial}).
    The value of kappa is fixed by $\langle E \rangle$ through
    Eq.~(\ref{eqn:kappa}).
  }
  \begin{tabular}{cccccc}
    \toprule
    ~~~~ $N$ ~~~
    &
    ~~~ $\langle E \rangle$ (GeV) ~~~
    &
    ~~~~ $a_1$ (fm$^{-1}$) ~~~
    &
    ~~~~ $a_2$ (fm$^{-2}$) ~~~
    &
    ~~~~ $a_3$ (fm$^{-2}$) ~~~
    &
    ~~~~ $a_4$ (fm$^{-2}$) ~~~
    \\
    \hline
    1 &
    -0.16300 &
    -0.3111 &
    --- &
    --- &
    --- \\
    2 &
    -0.16338 &
    -0.4307 &
    0.0823 &
    --- &
    --- \\
    3 &
    -0.16340 &
    -0.4688 &
    0.1356 &
    -0.0178 &
    --- \\
    4 &
    -0.16340 &
    -0.4803 &
    0.1598 &
    -0.0341 &
    0.0033 \\
    \bottomrule
  \end{tabular}
  \label{tab:variational}
\end{table}

\begin{figure}
  \includegraphics[width=\textwidth]{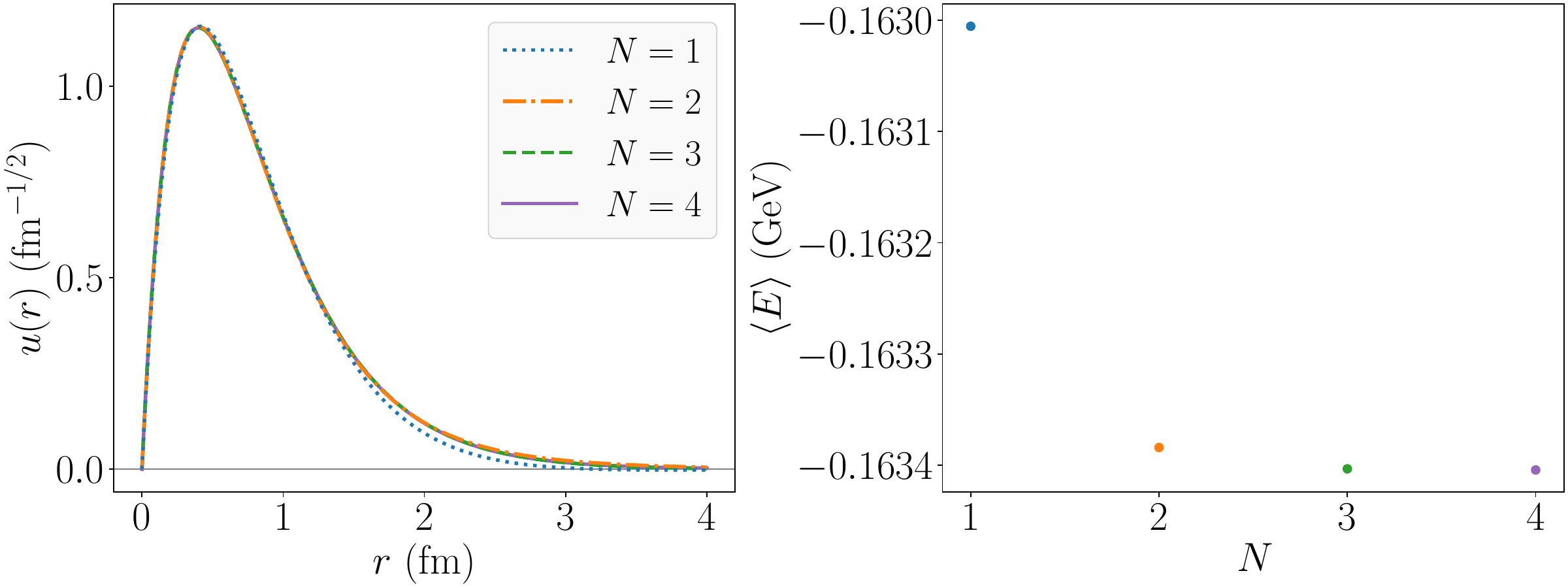}
  \caption{
    Estimates for the ground state
    wave function (left panel) and energy (right panel)
    for the Yukawa potential.
    The trial wave function of Eq.~(\ref{eqn:trial})
    has been scaled to be normalized to unity.
    Parameters are the same as in Tab.~\ref{tab:variational}.
  }
  \label{fig:variational}
\end{figure}

Results for the variational estimates of the Yukawa ground state
energy and wave function, with the trial form (\ref{eqn:trial})
and screening parameter $\mu=0.1~\mathrm{GeV}$,
are given in Tab.~\ref{tab:variational}.
Plots of $u(r)$ and the ground state energy are additionally given in
Fig.~\ref{fig:variational}.
Remarkable precision is already reached at $N=3$.

\begin{figure}
  \includegraphics[width=\textwidth]{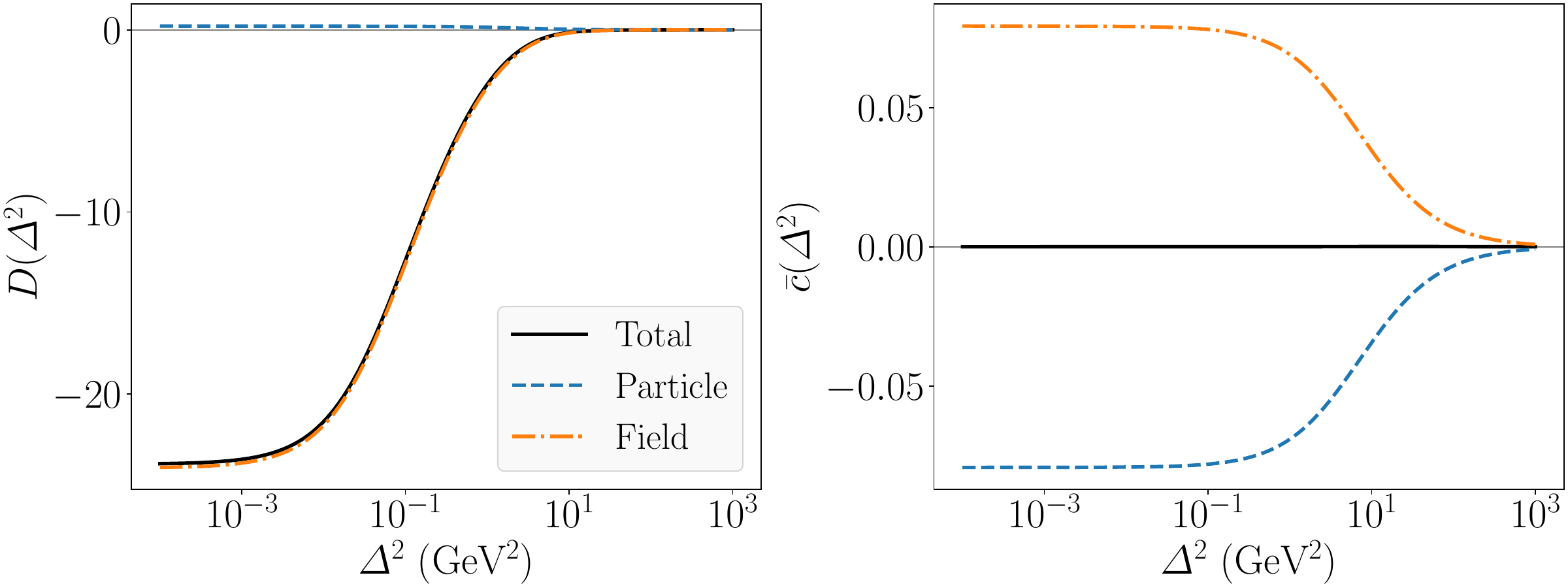}
  \caption{
    EMT form factors for the ground state of a Yukawa potential
    assuming the Yukawa field to have spin zero.
    The parameters used are $m_1 = m_2 = 1~\mathrm{GeV}$,
    $\mu = 0.1~\mathrm{GeV}$ and $\alpha = 1$.
    A variational method was used to estimate the wave function,
    with the trial form of Eq.~(\ref{eqn:trial}) and $N=3$.
    The lines have the same meaning as in Fig.~\ref{fig:massless}.
  }
  \label{fig:yukawa:zero}
\end{figure}

With a wave function in hand, we can obtain EMT form factors---using
Eqs.~(\ref{eqn:An}), (\ref{eqn:cn}) and (\ref{eqn:cn}) for particle contributions
and Eq.~(\ref{eqn:mff:force}) for the static field.
The field contributions depend on the spin of the field,
and for a first demonstration I will consider a spin-zero field.
I use the same parameters as in Tab.~\ref{tab:variational}
and wave function as in Fig.~\ref{fig:variational},
with $N=3$,
and assume equal charges $g_1 = g_2 = \sqrt{4\pi\alpha}$.
Since the field is spin-even,
$g_1$ and $g_2$ must have the same sign to produce an attractive potential.
The EMT form factors with these parameters are shown in
Fig.~\ref{fig:yukawa:zero}.

In the right panel of Fig.~\ref{fig:yukawa:zero},
the total $\bar{c}(\bd^2)$ vanishes---as
is required by local momentum conservation.
The numerical vanishing of $\bar{c}(\bd^2)$ is also a validation
that the wave function and force law are consistent with each other,
and accordingly provides a sanity check on the numerical estimate
of the wave function.
Most interestingly,
the left panel shows a large and negative $D(\bd^2)$
that is dominated by the static field contribution.
Like with the gravitationally bound system in
the bottom-left panel of Fig.~\ref{fig:massless},
$D(\bd^2)$ is negative because the field has even spin;
but unlike the gravitational system,
$D(0)$ is finite owing to the non-zero field mass $\mu$,
which exponentially screens the strength in the far field region.

\begin{figure}
  \includegraphics[width=\textwidth]{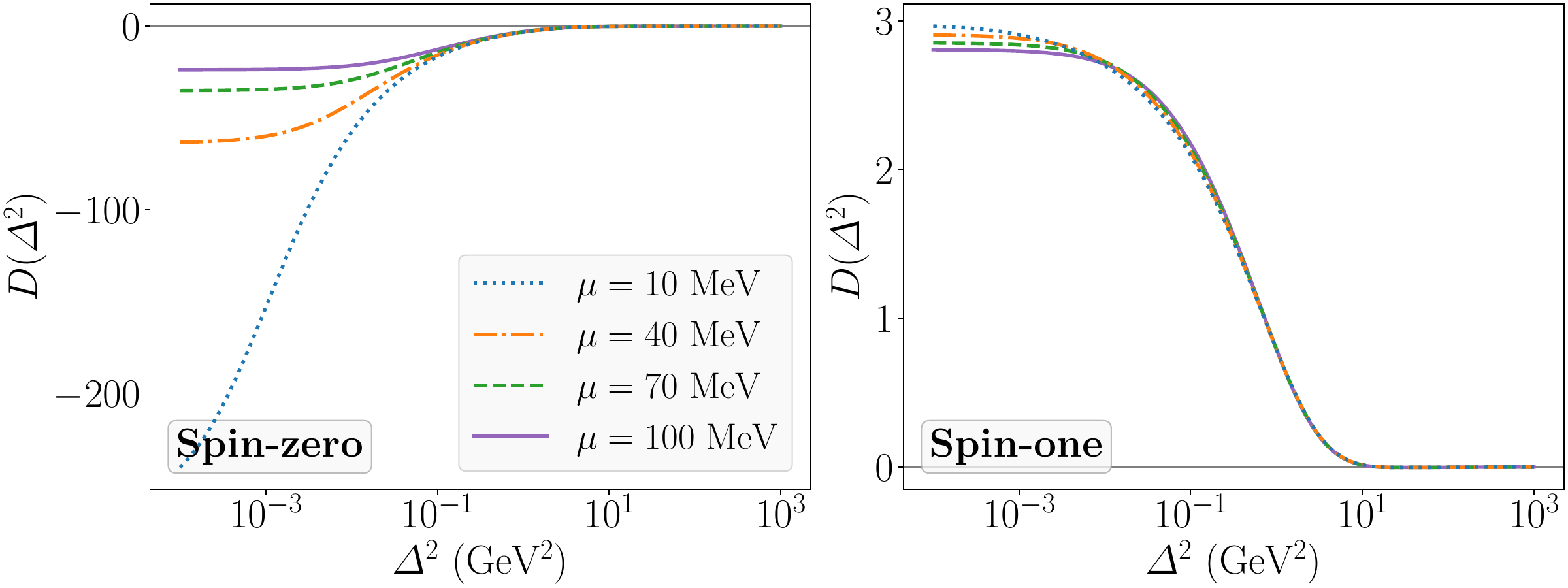}
  \caption{
    EMT form factors for the ground state of a Yukawa potential
    for both spin-zero (left panel) and spin-one (right panel) fields.
    The parameters $m_1 = m_2 = 1~\mathrm{GeV}$
    and $\alpha = 1$ are fixed, while $\mu$ varies.
  }
  \label{fig:yukawa:D}
\end{figure}

An especially interesting point of contrast is now to contrast the
$D(\bd^2)$ form factor
for solutions to the Yukawa potential assuming spin-zero and spin-one fields.
An attractive potential is chosen in both cases,
so $g_1 = g_2 = \sqrt{4\pi\alpha}$ for the spin-zero case
and $g_1 = -g_2 = \sqrt{4\pi\alpha}$ in the spin-one case.
In both cases, $m_1 = m_2 = 1~\mathrm{GeV}$ and $\alpha=1$.
I also consider, for both spins, a variety of field mass values:
$\mu \in \{10, 40, 70, 100\}~\mathrm{MeV}$.
The results are shown in Fig.~\ref{fig:yukawa:D}.

The left panel of Fig.~\ref{fig:yukawa:D}
shows the $D(\bd^2)$ form factor assuming a spin-zero field.
Because the field contribution dominates over the particle contribution,
the form factor is strictly negative.
Additionally---as expected from Eq.~(\ref{eqn:D0:massive})%
---the magnitude of $D(0)$ is (roughly)
inversely proportional to $\mu$.
In the massless limit, we should recover the massless result that
$D(0) = -\infty$, as seen in the gravitationally bound system.

The right panel of Fig.~\ref{fig:yukawa:D}
shows the $D(\bd^2)$ form factor assuming a spin-one field.
Much like the electrostatically bound system---see
the top-left panel of Fig.~\ref{fig:massless}---%
$D(\bd^2)$ is positive.
In contrast to the spin-zero case,
there is actually little dependence on the field mass,
with $D(0)$ hovering fairly close to the massless value of $3$.

There are several important lessons to take away from the contrast between
the panels of Fig.~\ref{fig:yukawa:D}---all of which echo the lessons
of the contrast between
the top and bottom panels of Fig.~\ref{fig:massless}.
The first lesson is that
the potential underdetermines the $D(\bd^2)$ form factor.
The potential and the wave function in the left and right panels
are identical, but the $D(\bd^2)$ form factor is not.

The second lesson is that the sign of $D(\bd^2)$ has nothing to do with
whether the forces operating are attractive or repulsive.
Exactly the same attractive force is operating in both the left
and right panels of Fig.~\ref{fig:yukawa:D},
but the $D(\bd^2)$ form factor couldn't look more different.

The third lesson---pointed out already in other prior
studies~\cite{Ji:2022exr,Freese:2024rkr,Ji:2025qax}%
---is that the sign of $D(0)$ has nothing to do with stability.
A system bound by spin-zero boson exchange is just as stable
as a system bound by spin-one boson exchange,
and yet their $D(0)$ have opposite signs.


\subsection{Multiple fields}

A last interesting case to consider is a bound system of two particles
interacting simultaneously through multiple fields.
As a concrete example,
I shall consider two particles bound by scalar boson exchange,
but which also interact electromagnetically and gravitationally.

Just as for the pure Yukawa potential,
the ground state wave function can be estimated using a variational method.
I follow the same procedure as in Sec.~\ref{sec:numeric:massive}---%
with the same trial wave function (\ref{eqn:trial}) and two-step procedure---%
but with the following potential:
\begin{align}
  \label{eqn:multifield}
  V(r)
  =
  -
  \frac{\alpha}{r}
  \e^{-\mu r}
  \pm
  \frac{\alpha_{\mathrm{QED}}}{r}
  -
  \frac{G m^2}{r}
  \,.
\end{align}
Here, $\alpha$ is the interaction strength for the scalar boson exchange.
The $\pm$ in the Coulomb interaction lets us consider both equal and opposite
charges.
I will again use $m = 1~\mathrm{GeV}$, $\mu = 0.1~\mathrm{GeV}$
and $\alpha=1$.

The contribution of each individual field to the EMT form factors
can be found by applying Eq.~(\ref{eqn:mff:force}).
For the Yukawa field contribution, for instance,
this formula is applied with the $\alpha$ and $\mu$ values given above;
whereas for the electrostatic field contribution,
the same formula is used,
but with $\alpha_{\mathrm{QED}}$ in place of $\alpha$
and $\mu$ set to zero.

\begin{figure}
  \includegraphics[width=\textwidth]{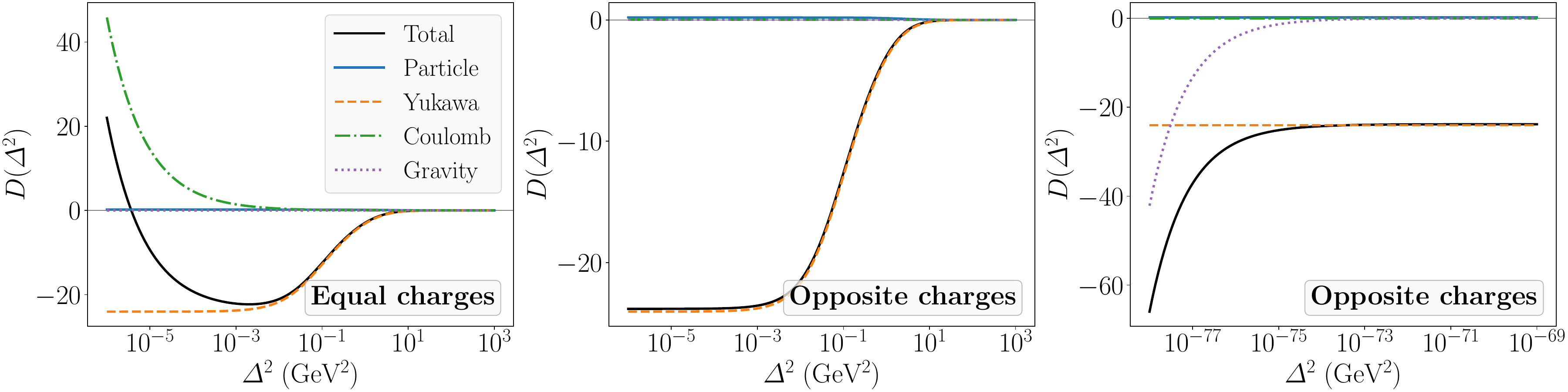}
  \caption{
    Breakdown of the $D(\bd^2)$ form factor
    for the ground state of the potential
    in Eq.~(\ref{eqn:multifield}),
    assuming the Yukawa part to arise from a spin-zero field.
    The left panel assumes the particles have equal electric charges,
    while the middle and right panels assume opposite charges.
    The middle and left panels differ in the $\bd^2$ range plotted.
    The particle contribution is non-zero,
    but too small to clearly discern on these plots.
    The parameters $m_1 = m_2 = 1~\mathrm{GeV}$,
    $\mu = 0.1~\mathrm{GeV}$ and $\alpha = 1$ are used.
  }
  \label{fig:multifield}
\end{figure}

Numerical results for the $D(\bd^2)$ form factor and its breakdown are
shown in Fig.~\ref{fig:multifield}.
Equal charges are considered in the left panel,
and opposite charges in the middle and right panels.
In the right panel, numerical evaluation of the exact formulas
for $D_\phi(\bd^2)$ become unstable,
so the asymptotic forms in Eqs.~(\ref{eqn:D0:massless}) and (\ref{eqn:D0:massive})
are used in this panel.

$D(0)$ is divergent in both cases.
For the equal-charge case (left panel),
$D(0) = +\infty$ because
the positive electrostatic contribution dominates over
the negative gravitational contribution.
The respective forms at small $\dl$ are:
\begin{align}
  D_{\mathrm{elec.}}(\bd^2)
  \approx
  \frac{\pi \alpha_{\mathrm{QED}} M}{\dl}
  \,,
  \qquad
  D_{\mathrm{grav.}}(\bd^2)
  \approx
  -
  \frac{\pi G m^2 M}{\dl}
  \,,
\end{align}
where
$
  \alpha_{\mathrm{QED}} \approx 0.00729735
  \gg
  G m^2 \approx 6.70883 \cdot 10^{-39}
$
for $m=1~\mathrm{GeV}$.
Although the equal-charge case has infinite $D(0)$,
the $D(\bd^2)$ form factor tends towards the Yukawa contribution
until slightly above $\bd^2 \sim 10^{-3}~\mathrm{GeV}^2$.
Thus, in an empirical setting,
one would not see the upturn to positive $D(\bd^2)$
nor the trend towards infinity unless probing an extreme forward
limit---one that is unlikely to be probed in any actual experiments.
This point has been discussed extensively by Varma and Schweitzer~\cite{Varma:2020crx},
and again by Mejia and Schweitzer~\cite{Mejia:2025oip},
in the context of classical models of the proton.
In fact, the left panel of Fig.~\ref{fig:multifield} looks remarkably
similar to the $D(\bd^2)$ results in Refs.~\cite{Varma:2020crx,Mejia:2025oip}.

On the other hand,
$D(0) = -\infty$ in the opposite charge case
(middle and right panels).
Since gravitation is so weak,
this divergence only becomes apparent when looking at absurdly small $\dl$.
In fact, to get
$
  D_{\mathrm{grav.}}(\bd^2)
  \approx
  1
$,
we need
$\dl \approx \pi G m^2 M$,
meaning $\bd^2 \sim 10^{-75}~\mathrm{GeV}^2$.
Thus, much like the equal-charge case,
the divergent trend in $D(\bd^2)$ at small $\dl$
will not be seen in an actual empirical setting.
And of course, the point is moot anyway if we stipulate%
---as general relativity does---%
that the gravitational field does not carry stress.


\section{Further discussion}
\label{sec:discuss}

The main thesis and conclusions of this work have already been presented.
However, several pertinent points merit further attention.
First, since EMT form factors encode spatial stress distributions,
it is worth looking at these, if only briefly.
Second, there is a connection between the sign of the D-term for a static
force field and the D-term of the associated field quantum.
Lastly, it is worth drawing attention to the fact that
quantum chromodynamics seems to violate the pattern discovered in this work,
and speculating why this might happen.


\subsection{Spatial stress distributions}

The relationship between form factors and densities in systems with Galilean
symmetry has been clarified in the convolution formalism,
first pioneered by Yang Li \textsl{et al.}~\cite{Li:2022ldb}.
In this framework, the quantum expectation value of a local operator
such as the stress tensor $\hat{T}^{ij}(x)$ is given between a convolution
of some internal densities with smearing functions,
the latter of which describe dispersion by the barycentric wave packet.
For the stress tensor of non-relativistic spin-zero systems,
the required convolution formula was given in Ref.~\cite{Freese:2024rkr}:
\begin{align}
  \label{eqn:convolution}
  \langle \Psi(t) | \hat{T}_a^{ij}(\bm{x}) | \Psi(t) \rangle
  =
  \int \d^3 R \,
  \left\{
    \left(
    -
    \Psi^*(\bm{R},t)
    \frac{\lrn^i \lrn^j}{4M}
    \Psi(\bm{R},t)
    \right)
    \mathfrak{a}_a(\bm{x} - \bm{R})
    +
    \mathfrak{t}_a^{ij}(\bm{x} - \bm{R})
    \Psi^*(\bm{R},t)
    \Psi(\bm{R},t)
    \right\}
  \,,
\end{align}
where:
\begin{align}
  \mathfrak{a}_a(\bm{b})
  &=
  \int \frac{\d^3\dl}{(2\pi)^3}
  A_a(\bd^2)
  \e^{- i\bd\cdot\bm{b}}
  \\
  \label{eqn:stress:fourier}
  \mathfrak{t}_a^{ij}(\bm{b})
  &=
  \int \frac{\d^3\dl}{(2\pi)^3}
  \left\{
    \frac{\dl^i \dl^j - \bd^2 \delta^{ij}}{4M}
    D_a(\bd^2)
    -
    M \delta^{ij} \bar{c}_a(\bd^2)
    \right\}
  \e^{- i\bd\cdot\bm{b}}
  \,.
\end{align}
An extension to the entire non-relativistic energy-momentum tensor
and an explanation for how such convolutions arise naturally in
de Broglie-Bohm pilot wave theory~\cite{db:pilot,Bohm:1951xw,Bohm:2006und}
can be found in Ref.~\cite{Freese:2025tqd}.
In brief, the first term in Eq.~(\ref{eqn:convolution}) describes how
dispersion of the barycentric wave packet $\Psi(\bm{R},t)$
produces dynamic or ram pressure,
while the second term describes how true internal stresses
$\mathfrak{t}^{ij}(\bm{b})$
are smeared out due to the uncertainty in the barycentric position.

As described by Polyakov and Schweitzer in their famous review~\cite{Polyakov:2018zvc},
the internal stress tensor can be further decomposed into an isotropic pressure
$p(b)$ and a pressure anisotropy $s(b)$,
the latter of which is often called the shear:
\begin{align}
  \mathfrak{t}^{ij}_a(\bm{b})
  =
  \delta^{ij}
  p(b)
  +
  Y_2^{ij}(\hat{b})
  S(b)
  \,.
\end{align}
The isotropic pressure in particular has been the focus of much attention.
When summed over constituents, it obeys the von Laue condition~\cite{Laue:1911emt}:
\begin{align}
  \int \d^3 b \,
  p(b)
  =
  0
  \,,
\end{align}
and the $\bm{b}^2$-weighted moment is related to the D-term:
\begin{align}
  D(0)
  =
  \frac{1}{3M}
  \int \d^3 b \,
  b^2 p(b)
  \,.
\end{align}
The sign of $D(0)$ thus tells us whether the long-distance region
is dominated by positive compressive pressure or negative tensile pressure.

\begin{figure}
  \includegraphics[width=\textwidth]{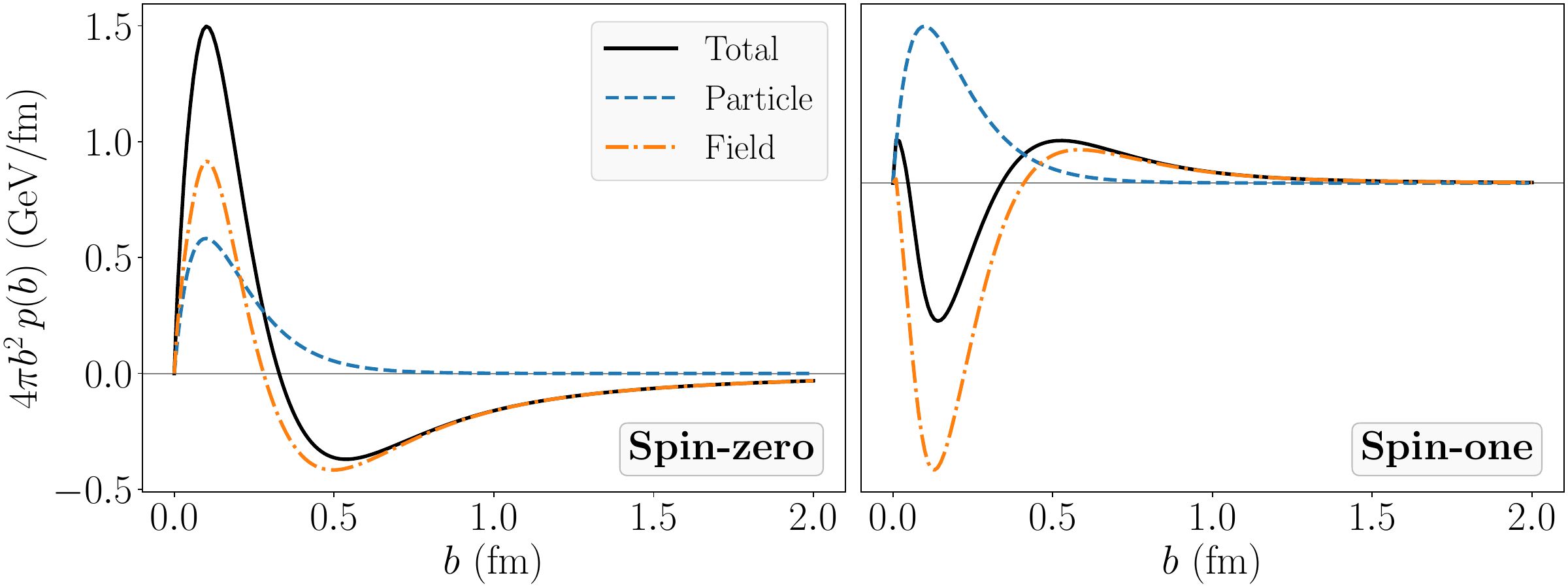}
  \caption{
    Isotropic pressure $p(b)$ of a two-particle system bound by a Yukawa potential
    with $m_1 = m_2 = 1~\mathrm{GeV}$,
    $\mu = 0.1~\mathrm{GeV}$ and $\alpha = 1$.
    The left and right panels differ only in whether the field responsible
    for the Yukawa potential is assumed to be spin-zero (left panel)
    or spin-one (right panel).
    Lines have the same meaning as Fig.~\ref{fig:massless}.
  }
  \label{fig:pressure}
\end{figure}

In the present context, plots of the particle and field
contributions to the isotropic pressure can reinforce the lessons
of the $D(\bd^2)$ plots in Fig.~\ref{fig:yukawa:D}.
The contribution of the $a$th constituent
(with $a\in\{1,2,\phi\}$)
to the isotropic pressure is given by the Bessel transform:
\begin{align}
  p_a(b)
  =
  -
  \frac{1}{2\pi^2}
  \int_0^\infty \d \dl \, \dl^2
  \left\{
    \frac{\dl^2}{6 M}
    D_a(\dl^2)
    +
    M
    \bar{c}_a(\dl^2)
    \right\}
  j_0(\dl b)
  \,.
\end{align}

I have calculated the particle and field contributions
to the isotropic pressure
for a system bound by a Yukawa potential
with the help of the hankel package~\cite{Murray2019},
and present the results in Fig.~\ref{fig:pressure}.
I again use $m_1 = m_2 = 1~\mathrm{GeV}$,
$\mu=0.1~\mathrm{GeV}$ and $\alpha=1$.
The left and right panels differ only in the spin of the
field responsible for the Yukawa potential.
I again stress that the potential and the wave function in both panels
are exactly identical,
but the form factor $D_\phi(\bd^2)$---and accordingly the
pressure profile---differ because of the spin of the field.
As suggested by the D-terms,
the system bound by a spin-zero field has compressive pressure
at short distances and tensile pressure at long distances,
while the system bound by a spin-one field has this ordering largely reversed
(aside from a positive-pressure core
owing to dominance by the particle pressure contribution at the center).
Since the potential is identical in both cases,
the ordering of the compressive and tensile regions
has no impact on stability.


\subsection{D-terms of associated quanta}

One curiosity worth drawing attention to is that the sign
of a static field's contribution to the D-term
correlates with the D-term of the field's quanta.
The quanta of elementary spin-zero fields are well-known to have
negative D-terms%
---for instance $D(0) = -1$ for the free Klein-Gordon field~\cite{Hudson:2017xug}
or $D(0)=-\frac{1}{3}$ for $\phi^4$ theory~\cite{Maynard:2024wyi}.
As we saw above, static abelian spin-zero fields also give negative contributions
to the D-term of a composite system.

The quanta of elementary spin-one fields
on the other hand
give positive D-terms,
with the photon infamously having $D(0) = 1$~\cite{Milton:1977je,Freese:2022ibw}.
At the same time, a static abelian spin-one field---such as the
electrostatic field---gives a positive contribution to
a composite system's D-term.

It's worth checking whether this pattern continues for spin-two fields
such as the gravitational field.
I must again raise the disclaimer that the standard formulation of
general relativity does not attribute an EMT to the gravitational field,
so the analysis to follow is necessarily rather speculative.
A variety of energy-momentum pseudotensors do exist in the literature%
---with the Landau-Lifshitz pseudotensor~\cite{landau1975classical}
being the most popular---%
but the energy-momentum pseudotensor is not unique.
The non-uniqueness problem persists in the regime of linearized gravity
that is relevant to describing gravitons;
see Refs.~\cite{Baker:2021qqi,Taylor:2024vvj}
for meticulous surveys of possible EMT choices
and their relationships to Noether's theorems.
For simplicity, I will adopt a normal-ordered version of
the gauge-dependent pseudotensor proposed by
Butcher, Hobson and Lasenby (BHL)~\cite{Butcher:2010ja}:
\begin{align}
  \label{eqn:BHL}
  T^{\mu\nu}
  =
  \frac{1}{32\pi G}
  :
  \left(
  (\partial^\mu h^{\alpha\beta})
  (\partial^\nu h_{\alpha\beta})
  -
  \frac{1}{2}
  (\partial^\mu h)
  (\partial^\nu h)
  -
  \frac{1}{2}
  \eta^{\mu\nu}
  (\partial^\rho h^{\alpha\beta})
  (\partial^\rho h_{\alpha\beta})
  +
  \frac{1}{4}
  \eta^{\mu\nu}
  (\partial^\rho h)
  (\partial^\rho h)
  \right)
  :
  \,,
\end{align}
where $\eta_{\alpha\beta} = \mathrm{diag}(+1,-1,-1,-1)$ is the flat Minkowski
spacetime metric,
$h_{\alpha\beta}(x) = g_{\alpha\beta}(x) - \eta_{\alpha\beta}$ is a deviation
of the metric from the flat background,
and $h(x) = \eta^{\alpha\beta} h_{\alpha\beta}(x)$.
This EMT is derived by BHL assuming the harmonic gauge:
\begin{align}
  \label{eqn:harmonic}
  \partial_\mu h^{\mu\nu}
  -
  \frac{1}{2}
  \partial^\nu h
  =
  0
  \,,
\end{align}
so its application requires a gravitational field in the harmonic gauge.

Next, we need the normal mode expansion for the gravitational field~\cite{Mehraeen:2026eaz}:
\begin{align}
  h^{\mu\nu}(x)
  =
  \sqrt{32\pi G}
  \sum_{\sigma=+,\times}
  \int \frac{\d^3k}{2|\bm{k}|(2\pi)^3}
  \varepsilon^{\mu\nu}(\bm{k},\sigma)
  \Big\{
    \e^{-ik\cdot x}
    a(\bm{k},\sigma)
    +
    \e^{+ik\cdot x}
    a^\dagger(\bm{k},\sigma)
    \Big\}
  \,.
\end{align}
The presence of a factor $\sqrt{32\pi G}$ is related to the appearance of
an overall factor
$\frac{1}{64\pi G}$
in the associated
linearized action,
and is required to ensure the conventional
covariant normalization for the creation and annihilation operators:
\begin{align}
  [ a(\bm{k}, \sigma), a^\dagger(\bm{k}',\sigma') ]
  =
  2|\bm{k}| (2\pi)^3 \delta^{(3)}(\bm{k} - \bm{k}')
  \,.
\end{align}
See Ref.~\cite{Mehraeen:2026eaz} for further details.
$\varepsilon^{\mu\nu}(\bm{k},\sigma)$ are the polarization tensors,
and the polarization modes are marked $+$ and $\times$ for the standard
plus and cross polarizations of gravitational waves~\cite{Flanagan:2005yc,Hsiang:2024qou}.
These tensors satisfy:
\begin{align}
  \label{eqn:satisfy}
  \begin{split}
    \varepsilon_{\mu\nu}(\bm{k},\sigma)
    &=
    \varepsilon_{\nu\mu}(\bm{k},\sigma)
    \\
    \varepsilon_{\mu\nu}(\bm{k},\sigma)
    \varepsilon^{\mu\nu}(\bm{k},\sigma')
    &=
    \delta_{\sigma\sigma'}
    \\
    \eta_{\mu\nu}
    \varepsilon^{\mu\nu}(\bm{k},\sigma)
    &=
    0
    \\
    k_\mu
    \varepsilon^{\mu\nu}(\bm{k},\sigma)
    =
    k_\mu
    \varepsilon^{\mu\nu}(\bm{k},\sigma)
    &=
    0
    \,.
  \end{split}
\end{align}
In words, they are symmetric, traceless and transverse.
(Explicit constructions for the graviton polarization tensors
are given in Appendix~\ref{sec:graviton}.)
Because they are traceless and transverse,
the graviton field automatically satisfies the harmonic gauge condition (\ref{eqn:harmonic})
and the BHL EMT (\ref{eqn:BHL}) can be used.

Sandwiching the BHL EMT and between
one-graviton momentum kets gives:
\begin{align}
  \langle \bm{p}', \lambda' |
  T^{\mu\nu}(0)
  | \bm{p}, \lambda \rangle \rangle
  =
  \left(
  2 P^\mu P^\nu
  -
  \frac{\dl^\mu \dl^\nu - \dl^2 \eta^{\mu\nu}}{2}
  \right)
  \varepsilon_{\mu\nu}(\bm{p},\lambda)
  \varepsilon^{\mu\nu}(\bm{p}',\lambda')
  \,.
\end{align}
Using the explicit polarization tensors in Appendix~\ref{sec:graviton},
one can show
\begin{align}
  \varepsilon_{\mu\nu}(\bm{p},\lambda)
  \varepsilon^{\mu\nu}(\bm{p}',\lambda')
  =
  \delta_{\lambda\lambda'}
\end{align}
and thus:
\begin{align}
  \langle \bm{p}', \lambda' |
  T^{\mu\nu}(0)
  | \bm{p}, \lambda \rangle \rangle
  =
  \left(
  2 P^\mu P^\nu
  -
  \frac{\dl^\mu \dl^\nu - \dl^2 \eta^{\mu\nu}}{2}
  \right)
  \delta_{\lambda\lambda'}
  \,.
\end{align}
Comparing to the standard EMT form factor breakdown of a spin-zero hadron,
\begin{align}
  \langle \bm{p}' | T^{\mu\nu}(0) | \bm{p} \rangle
  =
  2 P^\mu P^\nu
  A(\dl^2)
  +
  \frac{\dl^\mu \dl^\nu - \dl^2 \eta^{\mu\nu}}{2}
  D(\dl^2)
  \,,
\end{align}
suggests the following identifications for the EMT form factors
of a graviton:
\begin{align}
  \begin{split}
    A_{\mathrm{graviton}}(\dl^2)
    &=
    1
    \\
    D_{\mathrm{graviton}}(\dl^2)
    &=
    -1
    \,.
  \end{split}
\end{align}
The D-term of a graviton is apparently negative.
Since the static gravitational field gives a negative contribution to a bound
system's D-term,
this appears to validate the pattern.


\subsection{Anticipated differences in Yang-Mills theory}

The $D(\bd^2)$ form factor of many hadrons---crucially
including the proton and pion---appears to be consistently negative.
This finding is persistent throughout empirical
extractions~\cite{Kumano:2017lhr,Burkert:2018bqq,Kumericki:2019ddg,Duran:2022xag,CLAS:2026lls,CLAS:2026bis},
lattice QCD computations~\cite{Hackett:2023rif,Hackett:2023nkr},
and effective models of QCD~\cite{Mai:2012yc,Freese:2019bhb,Neubelt:2019sou,Lorce:2022cle,Mamo:2022eui,Cao:2023ohj}.
Many researchers have speculated that this negativity%
---or, at the very least, the negativity of $D(0)$---%
is a necessary stability
condition~\cite{Perevalova:2016dln,Polyakov:2018zvc,Lorce:2018egm}.
However, counter-examples of stable systems for which $D(0) > 0$ have been
found~\cite{Metz:2021lqv,Freese:2022ibw,Ji:2022exr,Freese:2024rkr},
and one of the principal findings of this work is that two-body quantum systems
with identical wave functions can have $D(\bd^2)$ with opposite signs.
In fact, systems bound by abelian spin-one fields naturally have positive
$D(\bd^2)$, and thus $D(0) > 0$---but are stable nevertheless.

The fact that hadrons typically have $D(0) < 0$ now appears \emph{atypical}.
Hadrons are bound by the spin-one gluon field, after all.
One crucial difference between the gluon field and electrostatic field, of course,
is that the gluon field is non-abelian.
The reason that hadrons have $D(0) < 0$
must somehow arise from the non-abelian nature of the gluon fields.

In fact, the hadronic $D(\bd^2)$ being negative may be a signature of
color flux confinement.
Just like the electrostatic field,
the unrenormalized chromoelectric field will give a positive-definite
contribution to the isotropic pressure.
Negative contributions can appear when calculating the self-field stresses
owing to the renormalization of short-distance divergences,
and physically these negative contributions correspond to Poincar\'e stresses
that hold a charged constituent together against disintegration
from self-repulsion.
For an electrically bound system,
the positive electrostatic field pressure is concentrated at greater distances
than the Poincar\'e stresses,
which are necessarily confined to the particles themselves.
Thus the $\bm{b}^2$ moment of the pressure and $D(0)$ are dominated by the positive
field pressure.

Chromoelectric fields, however,
are confined to thin flux tubes between color
charges~\cite{Maedan:1988yi,Maedan:1989ju,Baker:2024peg}.
We can thus reasonably expect the negative Poincar\'e stresses
to dominate at larger distances,
and thus for the $\bm{b}^2$ moment and $D(0)$ to be negative.
However, while a reasonable expectation,
this outcome is not immediately guaranteed.
The hypothesis that color flux confinement is responsible for the negative
hadronic $D(\bd^2)$ requires a detailed calculation within a consistent,
QCD-motivated model for the stress profile between color charges.
This is beyond the scope of the present work and
will be the subject of a follow-up study.


\section{Summary and outlook}
\label{sec:end}

In this work, I studied the contributions of static abelian force fields
to the EMT form factors of quantum two-body systems.
The principal finding is that the sign of the D-term form factor $D(\bd^2)$
depends on the spin of the field responsible for the binding force.
Spin-even fields give negative D-terms,
while spin-odd fields give positive D-terms.
Since the same potential energy function can be produced by forces mediated
by fields of either spin,
the D-term form factor is underdetermined by the potential energy function
and by the system's wave function.
Moreover, since systems with identical wave functions can be bound by forces
mediated by either even-spin or odd-spin fields,
the sign of the D-term is unrelated to the stability of the system.

There are crucial differences between hadrons and the simple two-body systems
considered in this work.
One of these is that hadrons are typically relativistic and field-theoretic
systems made up of an indefinite number of particles.
However, heavy quarkonia can reasonably be described as non-relativistic
two body systems bound by a potential energy function,
as indicated by the success of non-relativistic QCD~\cite{Brambilla:1999xf}.
Another crucial difference in hadrons---which equally applies to heavy
quarkonia---is that the gluon field responsible for mediating the color force
is non-abelian.

This makes the study of stress in heavy quarkonia a promising next step.
The present study found that the D-term for a two-body system bound
by an abelian spin-one field is positive,
but will a non-abelian field instead produce a negative D-term?
If so, then what exactly is responsible for this difference?
One possibility is that color flux confinement concentrates the positive,
compressive stresses naturally carried by the spin-one gluon field between
the quark and antiquark,
while the negative Poincar\'e stresses are concentrated at the particles themselves.
A detailed calculation is needed to judge the merit of this hypothesis.
Stresses in heavy quarkonia will accordingly be the main subject of a follow-up study.


\makeatletter{
\def\addcontentsline#1#2#3{}%
\def\tocsection#1#2#3{#3}%
\begin{acknowledgments}
  I warmly acknowledge helpful discussions with
  Peter Schweitzer and Alan Sosa.
  This material is based upon work supported by
  the Center for Nuclear Femtography,
  operated by the Southeastern Universities Research Association
  in Washington, D.C.\ under an appropriation from the Commonwealth of Virginia;
  and by the U.S.\ Department of Energy,
  Office of Science, Office of Nuclear Physics under Contract No.\ 89243126CSC000213.
  This work is dedicated to the memory of David A.\ Freese Jr.
\end{acknowledgments}
}
\makeatother


\makeatletter{
\def\addcontentsline#1#2#3{}%
\def\tocsection#1#2#3{#3}%
\section*{Data availability}

The open-source \textbf{deupack} repository~\cite{Freese_deupack_2026}
was used to generate the numerical data provided in this work.
}
\makeatother


\makeatletter{
\def\addcontentsline#1#2#3{}%
\def\tocsection#1#2#3{#3}%
\section*{Use of AI tools declaration}

No AI tools were used at any stage of this project.
}
\makeatother

\appendix


\section{Auxiliary function}
\label{sec:functions}

This appendix is dedicated to the auxiliary function
$\Phi(\zeta,\omega,\delta)$ defined in Eq.~(\ref{eqn:Phi:definition}).
The $y$ integral can be performed analytically,
but the resulting formula (\ref{eqn:Phi:result}) is not pretty.

Let's proceed with the derivation.
As a first step, we define some auxiliary variables:
\begin{align}
  u
  =
  \frac{y}{\sqrt{1+\omega}}
  \,, \qquad
  x
  =
  \zeta \sqrt{1+\omega}
  \,, \qquad
  w
  =
  \frac{\delta}{\sqrt{1+\omega}}
\end{align}
so that
\begin{align}
  y\zeta
  =
  xu
  \,, \qquad
  \zeta \sqrt{1-y^2 + \omega}
  =
  x \sqrt{1-u^2}
  \,, \qquad
  (y+\delta)\zeta
  =
  (u+w)x
  \,.
\end{align}
This allows Eq.~(\ref{eqn:Phi:definition}) to be rewritten:
\begin{align}
  \Phi(\zeta,\omega,\delta)
  =
  \int_{-1/\sqrt{1+\omega}}^{1/\sqrt{1+\omega}}
  \d u \,
  \frac{\e^{-x\sqrt{1-u^2}}}{\sqrt{1-u^2}}
  j_0\big((u+w)x\big)
  \,.
\end{align}
The next step is to use something like
\begin{align}
  j_0\big((u+w)x\big)
  =
  -
  \frac{1}{x}
  \mathrm{Im}
  \frac{\e^{-i(u+w)x}}{u+w}
  \,,
\end{align}
but this needs to be approached carefully,
since the real part of the integral
\begin{align*}
  \int_{-1/\sqrt{1+\omega}}^{1/\sqrt{1+\omega}}
  \d u \,
  \frac{
    \e^{-x\big(\sqrt{1-u^2} + i(u+w) \big)}
  }{(u+w)\sqrt{1-u^2}}
\end{align*}
diverges because of the pole at $u=-w$.
The pole can be avoided by adding a small imaginary part to $w$,
but doing so modifies the imaginary part of the integrand by adding
or subtracting a delta function:
\begin{align}
  -
  \mathrm{Im}
  \frac{\e^{-i(u+w)x}}{(u+w \pm i\epsilon)}
  =
  \frac{\sin\big((u+w)x\big)}{u+w}
  \pm
  \pi
  \delta(u+w)
  \,.
\end{align}
Fortunately, this modification can be cancelled out
by averaging the offset of $w$ by both $\pm i\epsilon$:
\begin{align}
  \label{eqn:reg:avg}
  j_0\big((u+w)x\big)
  =
  \frac{\sin\big((u+w)x\big)}{(u+w)x}
  =
  -
  \frac{1}{2x}
  \mathrm{Im}
  \left\{
    \frac{\e^{-i(u+w)x}}{u+w+i\epsilon}
    +
    \frac{\e^{-i(u+w)x}}{u+w-i\epsilon}
    \right\}
  \,,
\end{align}
which effectively amounts to using the principal value prescription
for the real part.
The auxiliary function can now be written:
\begin{align}
  \Phi(\zeta,\omega,\delta)
  =
  -
  \frac{1}{2x}
  \mathrm{Im}
  \e^{-iwx}
  \int_{-1/\sqrt{1+\omega}}^{1/\sqrt{1+\omega}}
  \d u \,
  \frac{
    \e^{-x\big(\sqrt{1-u^2} + iu \big)}
  }{\sqrt{1-u^2}}
  \left\{
    \frac{1}{u+w+i\epsilon}
    +
    \frac{1}{u+w-i\epsilon}
    \right\}
  \,.
\end{align}
Although the imaginary part of the integral is being taken,
it is important to ensure that the real part is finite and well-defined;
it is not mathematically meaningful to take the imaginary part of
an infinite result.
We will also soon see that the regularization has an impact on the final result.

\begin{figure}
  \includegraphics[scale=1]{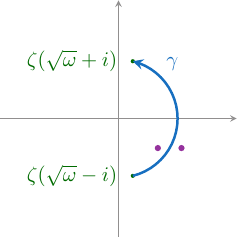}
  ~~~~
  \includegraphics[scale=1]{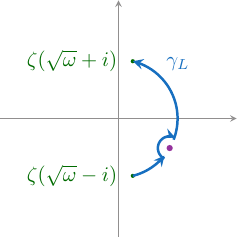}
  ~~~~
  \includegraphics[scale=1]{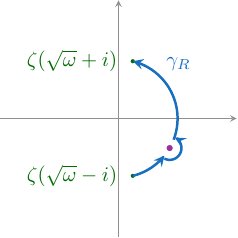}
  \caption{
    Paths used in the integrals in Eqs.~(\ref{eqn:gamma})
    [left panel]
    and (\ref{eqn:gamma:LR})
    [all three panels].
    The start and end values of $z$ are written in green text,
    and the important poles
    [$r_1^\pm$ in Eq.~(\ref{eqn:poles})]
    are depicted as purple dots.
    (The irrelevant poles
    [$r_2^\pm$ in Eq.~(\ref{eqn:poles})]
    are not depicted, and fall on the left half of $\mathbb{C}$.)
    In Eq.~(\ref{eqn:gamma}),
    there are split poles on either side of the contour,
    while in Eq.~(\ref{eqn:gamma:LR}),
    the path is deformed and the poles moved to
    $z = \sqrt{1-w^2} - iw$.
  }
  \label{fig:contours}
\end{figure}

The next step is another change of variables,
this time to the complex variable
\begin{align}
  z
  =
  x
  \big(
  \sqrt{1-u^2}
  +
  i u
  \big)
  \,, \qquad
  \d z
  =
  \frac{i z}{\sqrt{1-u^2}} \d u
  \,,
\end{align}
which inverts to
\begin{align}
  u
  =
  \frac{z^2 - x^2}{2i x z}
  \,.
\end{align}
This allows $\Phi(\zeta,\omega,\delta)$ to be rewritten:
\begin{align}
  \label{eqn:gamma}
  \Phi(\zeta,\omega,\delta)
  =
  -
  \mathrm{Im}
  \e^{-i x w}
  \int_\gamma \d z \,
  \left\{
    \frac{
      \e^{-z}
    }{
      z^2 - x^2 + 2i x (w+i\epsilon) z
    }
    +
    \frac{
      \e^{-z}
    }{
      z^2 - x^2 + 2i x (w-i\epsilon) z
    }
    \right\}
\end{align}
where the path $\gamma$ through the complex plane is depicted
in the left panel of Fig.~\ref{fig:contours}.
The roots of the denominators in this integrand are:
\begin{align}
  \label{eqn:poles}
  \begin{split}
    r_1^{\pm}
    &=
    x
    \big( \sqrt{1-w^2} - i w \big)
    \pm
    x(1-iw) \epsilon
    +
    \mathcal{O}(\epsilon^2)
    \equiv
    r_1
    \pm
    x(1-iw) \epsilon
    \\
    r_2^{\pm}
    &=
    -
    x
    \big( \sqrt{1-w^2} + i w \big) \pm
    +
    \mathcal{O}(\epsilon)
    \equiv
    r_2
    \,.
  \end{split}
\end{align}
The pole at $u=w$ in the unregulated integral
appears in the right half of $\mathbb{C}$ in terms of $z$,
namely at $z=\sqrt{1-w^2}-iw$.
This corresponds to the $r_1^\pm$ root.
The $\epsilon$ regulation prescription is not necessary
for $r_2$, which is why I dropped it.
As for the $r_1^\pm$ poles,
these are to the right ($+$) and left ($-$)
of the integration path $\gamma$.
The $r_1^\pm$ poles can be migrated to $r_1$
(setting $\epsilon=0$),
if the integration path is deformed to deflect to the left ($+$)
or right ($-$) of the pole;
see the middle and right panels of Fig.~\ref{fig:contours}.
The auxiliary function can then be written:
\begin{align}
  \label{eqn:gamma:LR}
  \Phi(\zeta,\omega,\delta)
  =
  -
  \frac{1}{x\sqrt{1-w^2}}
  \mathrm{Im}
  \,
  \e^{-i x w}
  \left\{
    \frac{1}{2}
    \left(
    \int_{\gamma_L} \d z \,
    +
    \int_{\gamma_R} \d z \,
    \right)
    \frac{ \e^{-z} }{ z - r_1 }
    -
    \int_{\gamma} \d z \,
    \frac{ \e^{-z} }{ z - r_2 }
    \right\}
  \,.
\end{align}

\begin{figure}
  \includegraphics[scale=1]{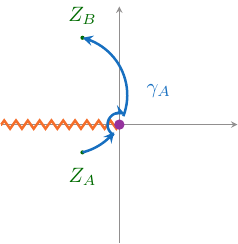}
  ~~~~
  \includegraphics[scale=1]{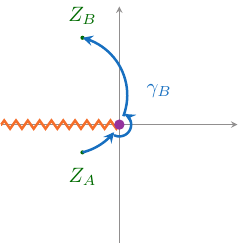}
  ~~~~
  \includegraphics[scale=1]{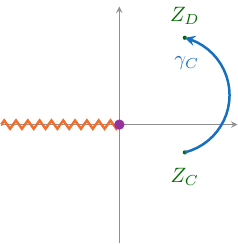}
  \caption{
    The integration paths used in Eq.~(\ref{eqn:gamma:ABC}).
    These paths are obtained by shifting the paths in
    Fig.~\ref{fig:contours}
    to the left for $\gamma_A$ and $\gamma_B$
    [left and middle panels],
    and to the right for $\gamma_C$ [right panel].
    The endpoint values are written in green text.
    The integrands in Eq.~(\ref{eqn:gamma:ABC})
    have a pole at the origin, depicted by a large purple dot.
    The integrals over the paths $\gamma_B$ and $\gamma_C$
    can be related to the exponential integral function
    $E_1(z)$,
    since these paths do not cross the negative real axis
    (depicted as an orange zigzag line).
  }
  \label{fig:contours2}
\end{figure}

The next step is another variable substitution,
either $z \rightarrow z+r_1$ or $z \rightarrow z+r_2$.
These give:
\begin{align}
  \label{eqn:gamma:ABC}
  \Phi(\zeta,\omega,\delta)
  =
  -
  \frac{1}{x\sqrt{1-w^2}}
  \,
  \mathrm{Im}
  \e^{-i x w}
  \left\{
    \frac{1}{2}
    \e^{-r_1}
    \left(
    \int_{\gamma_A} \d z \,
    +
    \int_{\gamma_B} \d z \,
    \right)
    \frac{ \e^{-z} }{ z }
    -
    \e^{-r_2}
    \int_{\gamma_C} \d z \,
    \frac{ \e^{-z} }{ z }
    \right\}
  \,,
\end{align}
where the contours $\gamma_A$, $\gamma_B$ and $\gamma_C$
are given in Fig.~\ref{fig:contours2},
and where the endpoints of the paths are:
\begin{align}
  \begin{split}
    Z_A
    &=
    \zeta
    \big[
      \sqrt{\omega} - i
      - \sqrt{1+\omega-\delta^2} + i \delta
      \big]
    \\
    Z_B
    &=
    \zeta
    \big[
      \sqrt{\omega} + i
      - \sqrt{1+\omega-\delta^2} + i \delta
      \big]
    \\
    Z_C
    &=
    \zeta
    \big[
      \sqrt{\omega} - i
      + \sqrt{1+\omega-\delta^2} + i \delta
      \big]
    \\
    Z_D
    &=
    \zeta
    \big[
      \sqrt{\omega} + i
      + \sqrt{1+\omega-\delta^2} + i \delta
      \big]
    \,.
  \end{split}
\end{align}
For the integrals over $\gamma_B$ and $\gamma_C$,
we can immediately write the results in terms of the exponential integral
function $E_1(z)$, defined on the principal sheet as~\cite{NIST:DLMF}:
\begin{align}
  E_1(z)
  =
  \int_z^\infty \d t \frac{\e^{-t}}{t}
  \,, \qquad
  \mathrm{arg}(z) \neq \pi
  \,,
\end{align}
where the path from $z$ to $\infty$ does not cross the branch cut
along the negative real axis.
The paths $\gamma_B$ and $\gamma_C$ avoid the negative real axis,
so can immediately be written in terms of $E_1$:
\begin{align}
  \begin{split}
    \int_{\gamma_B} \d z \,
    \frac{ \e^{-z} }{ z }
    &=
    E_1(Z_A)
    -
    E_1(Z_B)
    \\
    \int_{\gamma_C} \d z \,
    \frac{ \e^{-z} }{ z }
    &=
    E_1(Z_C)
    -
    E_1(Z_D)
    \,.
  \end{split}
\end{align}
Since the integral over $\gamma_A$ passes through the negative real axis
(and cannot be deformed away from this because of the pole at the origin),
an analytic continuation of $E_1(z)$ onto secondary sheets is necessary.
To this end, the identity~\cite{NIST:DLMF}
\begin{align}
  E_1\big( z \e^{2\pi n i} \big)
  =
  E_1(z)
  -
  2\pi n i
\end{align}
is helpful.
This tells us:
\begin{align}
  \int_{\gamma_A} \d z \,
  \frac{ \e^{-z} }{ z }
  &=
  E_1(Z_A)
  -
  E_1(Z_B)
  -
  2\pi i
  \,.
\end{align}
This extra $-2\pi i$ survives the $\mathrm{Im}$ operation,
and contributes to the final result.
It is halved---see the factor $\frac{1}{2}$ in Eq.~(\ref{eqn:gamma:ABC})---%
because the regulation prescription adopted back in
Eq.~(\ref{eqn:reg:avg}) averaged over offsets by $\pm i\epsilon$,
only one of us leads us to a secondary sheet of the complex numbers.
It was important that the regularization at that stage was done correctly,
because otherwise the final result would be missing a term.

Besides plugging in the integration results,
a few more manipulations can be done to make
Eq.~(\ref{eqn:gamma:ABC}) a bit nicer.
The imaginary parts of $r_1$ and $r_2$ are both $-ixw$,
which can be used to eliminate the factor $\e^{-ixw}$.
We can rewrite $x\sqrt{1-w^2} = \zeta\sqrt{1+\omega-\delta^2}$.
With these changes, we get:
\begin{align}
  \label{eqn:Phi:result}
  \Phi(\zeta,\omega,\delta)
  =
  \frac{1}{\zeta\sqrt{1+\omega-\delta^2}}
  \left\{
    \e^{-\zeta\sqrt{1+\omega-\delta^2}}
    \Big(
    \mathrm{Im}\big[
      E_1(Z_B)
      -
      E_1(Z_A)
      \big]
    +
    \pi
    \Big)
    +
    \e^{+\zeta\sqrt{1+\omega-\delta^2}}
    \mathrm{Im}\big[
      E_1(Z_C)
      -
      E_1(Z_D)
      \big]
    \right\}
  \,.
\end{align}
I have unfortunately not been able to find a simpler expression.

While Eq.~(\ref{eqn:Phi:result}) is not pretty, it is usable.
SciPy's implementation of $E_1(z)$ can take complex arguments,
and for moderate values of $\zeta\sqrt{1+\omega-\delta^2}$,
Eq.~(\ref{eqn:Phi:result}) works quite well.
The formula fails, however, when
$\zeta\sqrt{1+\omega-\delta^2} \gtrsim 700$
owing to overflow in $\e^{+\zeta\sqrt{1+\omega-\delta^2}}$.
This problem can be mitigated by using the asymptotic expansion for $E_1(z)$:
\begin{align}
  E_1(z)
  \approx
  \frac{\e^{-z}}{z}
  \sum_{n=0}^N
  \frac{(-1)^n n!}{z^n}
  \,,
\end{align}
which holds at large $|z|$~\cite{NIST:DLMF}.
For the $E_1$ functions multiplying $\e^{+\zeta\sqrt{1+\omega-\delta^2}}$
in Eq.~(\ref{eqn:Phi:result}), this means:
\begin{align}
  \e^{+\zeta\sqrt{1+\omega-\delta^2}}
  E_1(Z_{C,D})
  \approx
  \frac{
    \exp\left\{\big[-\sqrt{\omega}+i(-\delta\pm1)\big]\zeta\right\}
  }{
    Z_{C,D}
  }
  \sum_{n=0}^N
  \frac{
    (-1)^n n!
  }{
    (Z_{C,D})^n
  }
  \,,
\end{align}
with the $+$ in $\pm$ corresponding to $Z_C$ and the $-$ to $Z_D$.
Overflow errors can be avoided by
using this asymptotic form whenever $\zeta\sqrt{1+\omega}$ is too large.
Reasonable precision is reached by using the asymptotic form
with $N=3$ whenever $\zeta\sqrt{1+\omega-\delta^2} \geq 50$.

\begin{figure}
  \includegraphics[width=\textwidth]{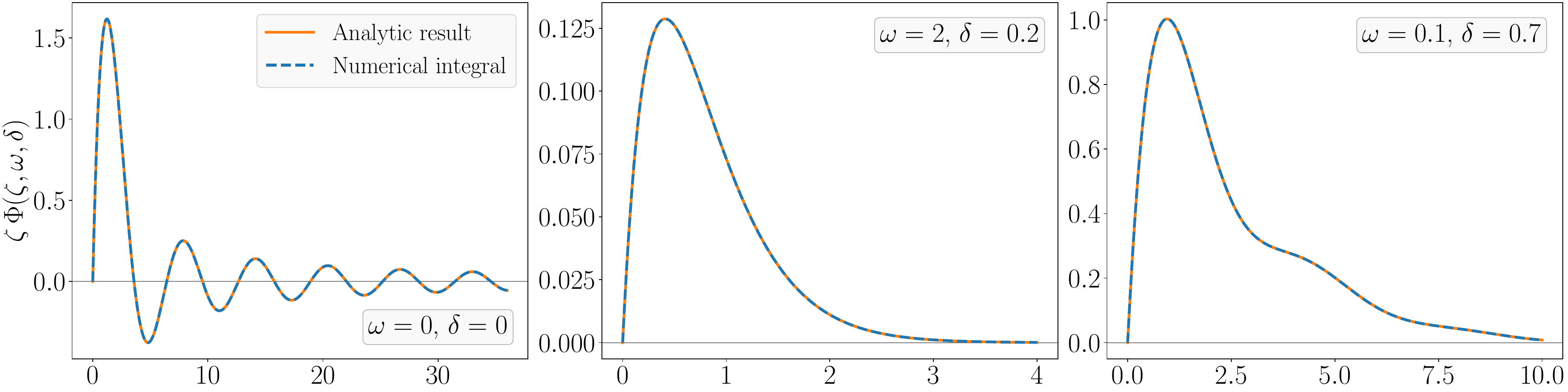}
  \caption{
    Plots of $\zeta \, \Phi(\zeta,\omega,\delta)$
    against $\zeta$ for three values of $(\omega,\delta)$.
    In each panel,
    the solid orange curve depicts the right-hand side of
    Eq.~(\ref{eqn:Phi:result}),
    evaluated using SciPy's implementation of the complex $E_1$ function.
    The dashed blue curve depicts a numerical evaluation of the integral in
    Eq.~(\ref{eqn:Phi:definition}) using \texttt{quad\_vec}
    from SciPy's integration library.
    The analytic and numerical curves coincide exactly.
  }
  \label{fig:Phi}
\end{figure}

As a sanity check,
numerically evaluating the integral in Eq.~(\ref{eqn:Phi:definition})
gives perfect agreement with Eq.~(\ref{eqn:Phi:result}).
This is shown in Fig.~\ref{fig:Phi}.


\section{Explicit graviton polarization tensors}
\label{sec:graviton}

This appendix provides explicit polarization tensors for gravitons with
initial and final momenta
\begin{align}
  \begin{split}
    \bm{p}
    &=
    \bm{P}
    -
    \frac{1}{2} \bd
    \\
    \bm{p}'
    &=
    \bm{P}
    +
    \frac{1}{2} \bd
  \end{split}
\end{align}
respectively.
Primed quantities signify the final state, and unprimed the initial state:
\begin{align}
  \begin{split}
    \varepsilon_\lambda^{\mu\nu}
    &\equiv
    \varepsilon^{\mu\nu}(\bm{p},\lambda)
    \\
    \varepsilon_\lambda'^{\mu\nu}
    &\equiv
    \varepsilon^{\mu\nu}(\bm{p}',\lambda')
    \,.
  \end{split}
\end{align}

Firstly, the plus and cross polarizations can be written
in terms of symmetric and antisymmetric combinations of basis vectors~\cite{Hsiang:2024qou}:
\begin{align}
  \begin{split}
    \varepsilon_+^{\mu\nu}
    &=
    \frac{1}{\sqrt{2}}
    \Big( e_1^\mu e_2^\nu - e_1^\nu e_2^\mu \Big)
    \\
    \varepsilon_\times^{\mu\nu}
    &=
    \frac{1}{\sqrt{2}}
    \Big( e_1^\mu e_2^\nu + e_1^\nu e_2^\mu \Big)
    \,,
  \end{split}
\end{align}
and likewise for primed quantities.
The basis vectors themselves must obey orthonormality and transversity conditions:
\begin{align}
  \label{eqn:satisfy:2}
  \begin{split}
    e_1^2
    &=
    e_2^2
    =
    -1
    \\
    e_1 \cdot e_2
    &=
    0
    \\
    p\cdot e_1
    &=
    p\cdot e_2
    =
    0
    \,,
  \end{split}
\end{align}
and again likewise for primed quantities.
These conditions on the basis vectors ensure the polarization tensors satisfy
all of the conditions in Eq.~(\ref{eqn:satisfy}).

Since $p$ and $p'$ are lightlike, it is impossible to build a linear combination
of them that satisfies all of Eqs.~(\ref{eqn:satisfy:2}).
A third independent four-vector, $n$, must be introduced.
It is convenient to choose this four-vector to also be lightlike.
The required basis vectors can then be found by specializing those of
Refs.~\cite{Berger:2001zb,Cano:2003ju,Cosyn:2018rdm}
to a massless target.
Doing so gives:
\begin{align}
  \begin{split}
    e_1^\mu
    &=
    -
    \frac{1}{\sqrt{(1-\xi^2)}}
    \frac{1}{|\dl|}
    \left(
    (1+\xi) p'^\mu
    -
    (1-\xi) p^\mu
    +
    \frac{\dl^2 n^\mu}{2P^+}
    \right)
    \\
    e_1'^\mu
    &=
    -
    \frac{1}{\sqrt{(1-\xi^2)}}
    \frac{1}{|\dl|}
    \left(
    (1+\xi) p'^\mu
    -
    (1-\xi) p^\mu
    -
    \frac{\dl^2 n^\mu}{2P^+}
    \right)
    \\
    e_2^\mu
    &=
    e_2'^\mu
    =
    \frac{1}{\sqrt{(1-\xi^2)}}
    \frac{1}{|\dl|}
    \frac{\epsilon^{\mu p' p n}}{P^+}
    \,,
  \end{split}
\end{align}
where contraction with $n$ define the plus direction
(e.g., $P^+ = P\cdot n$), and where
\begin{align}
  \xi
  =
  -
  \frac{\dl^+}{2P^+}
  =
  -
  \frac{\dl\cdot n}{2(P\cdot n)}
\end{align}
is the skewness variable,
relevant to the physics of generalized parton distributions.
The kinematic identities
\begin{align}
  \begin{split}
    p^2
    &=
    p'^2
    =
    n^2
    =
    0
    \\
    p\cdot p'
    &=
    -
    \frac{\dl^2}{2}
  \end{split}
\end{align}
can be helpful in proving that these four-vectors indeed satisfy
the required conditions (\ref{eqn:satisfy:2}).

In addition to the required conditions, we also have:
\begin{align}
  e_1\cdot e_1'
  =
  -1
  \,.
\end{align}
Since $e_2 = e_2'$, we also have
\begin{align}
  e_1\cdot e_2'
  =
  e_2\cdot e_1'
  =
  0
  \,.
\end{align}
These conditions together entail:
\begin{align}
  \varepsilon_{\mu\nu}(\bm{p},\lambda)
  \cdot
  \varepsilon^{\mu\nu}(\bm{p}',\lambda')
  =
  \delta_{\lambda\lambda'}
  \,.
\end{align}


\bibliography{references.bib}

\end{document}